\documentclass[epj]{svjour}
\usepackage[mathscr]{eucal}
\usepackage{amssymb}
\usepackage{amsfonts}
\usepackage{ae}

\usepackage{dcolumn}
\usepackage{amsmath}
\usepackage{graphics,graphicx,epsfig}
\usepackage{hyperref}
\usepackage{epstopdf}

\usepackage{color}
\usepackage{ulem}

\begin{document}



\def\BE{\begin{equation}}
\def\EE{\end{equation}}
\def\BY{\begin{eqnarray}}
\def\BEA{\begin{eqnarray}}
\def\EY{\end{eqnarray}}
\def\EEA{\end{eqnarray}}
\def\L{\label}
\def\nn{\nonumber}
\def\({\left (}
\def\){\right)}
\def\<{\langle}
\def\>{\rangle}
\def\[{\left [}
\def\]{\right]}
\def\BA{\begin{array}}
\def\EA{\end{array}}
\def\ds{\displaystyle}
\def\c{^\prime}
\def\cc{^{\prime\prime}}

\title{High speed spatially multimode $\Lambda$-type atomic memory with arbitrary frequency detuning }
\author{T.~Golubeva\inst{1}, Yu.~Golubev\inst{1} \and O. Mishina\inst{2,3}, A. Bramati\inst{2}, J. Laurat\inst{2}, E. Giacobino\inst{2}}
\institute{St.~Petersburg State University, 198504 St.~Petersburg, Stary Petershof, ul. Ul'yanovskaya, 1, Russia \and Laboratoire Kastler Brossel,
Universit\'{e} Pierre et Marie Curie, Ecole Normale Sup\'{e}rieure, CNRS, Case 74, 4 place Jussieu, 75252 Paris Cedex 05, France \and Theoretische
Physik, Universit\"{a}t des Saarlandes, D-66123 Saarbr\"{u}cken, Germany}

\date{\today}

\abstract{We present a general model for an atomic memory using ultra-short pulses of light, which allows both spatial and temporal multimode
storage. The process involves the storage of a faint quantum light pulse into the spin coherence of the ground state of $\Lambda$-type 3-level atoms,
in the presence of a strong driving pulse. Our model gives a full description of the evolution of the field and of the atomic coherence in space and
time throughout the writing and the read-out processes. It is valid for any frequency detuning, from the resonant case to the Raman case, and allows
a detailed optimization of the memory efficiency.}

%
\authorrunning{T.~Golubeva et.al.}
\titlerunning{High speed $\Lambda$-type memory with arbitrary frequency detuning}

\maketitle

\section{Introduction}

For quantum telecommunications and for quantum information
processing, memory registers able to store quantum information
without measuring it are essential devices. A quantum memory relies
on an efficient coupling between light and matter, allowing
reversible mapping of quantum photonic information in and out of the
material system.  In the past years, several protocols have been
developed theoretically and
experimentally\cite{Lvovsky2009rev,Simon2010rev}. Storage and
retrieval of some of the basic states of light for quantum
communication such as a polarization q-bit \cite{Choi2008}, squeezed
light \cite{Honda2008,Appel2008} and entangled photons
\cite{Clausen2011,Saglamyurek2011} or faint coherent pulses at the
level of one to few photons have been realized
\cite{Cviklinski2008,Gisin2008,Sellars2010,Lam2011,Walmsley2011}.
Recent experiments have achieved large efficiencies
\cite{Sellars2010,Lam2011}.

However, the processing speed and the available bandwidth of the
memories remain a challenge for quantum memories. The first quantum
memory registers proposed more than a decade  ago \cite{Lukin2000}
involve the transfer of quantum information from light to atoms
(writing) and back from atoms to light (retrieval), using
electromagnetically induced transparency (EIT) in atomic three-level
transitions, and this process implies a limited bandwidth. The
storage protocol relies on a strong control field, generating EIT
for the weak field that carries the quantum signal to be stored. The
group velocity for the signal field is strongly reduced and the
signal pulse is compressed by several orders of magnitude. A signal
pulse can thus be contained inside the atomic medium, and before it
propagates outside the medium, the control is switched off. The
quantum variables of the signal field are then converted from a
purely photonic state to a collective spin coherence. For read-out,
the control field is turned on again and the medium emits a weak
pulse, carrying the quantum information contained in the original
pulse. While in principle this allows direct mapping of the quantum
state of light into long lived coherences in the atomic ground
state, the bandwidth of the stored signal is strongly limited by the
transparency window associated to EIT.

Various methods have been proposed to achieve broadband memories and
escape the limitations linked to EIT. Very interesting protocols are
based on the implementation of controlled broadening, such as CRIB
(controlled reversible inhomogeneous broadening) \cite{Hetet2008},
using photon echo-type reversal \cite{Moiseev2001} and AFC (atomic
frequency comb) where an absorbing comb structure is created in the
medium \cite{Afzelius2009}. These techniques have been successfully
applied for echo-type light storage in rare-earth doped crystals
\cite{Tillel2010} and atomic vapours \cite{Hosseini2011}. These
methods allow broad bandwidth but they imply writing times which are
still rather long (of the order of microseconds). In the spatial
domain, an interesting phenomena, quantum holography, has been
proposed to implement 3D-memories \cite{Sokolov2008,Sokolov2010}.

An alternative method is based on the use of a broadband, ultrafast control and signal field pulses for the writing process. It relies on two
different atomic transitions sharing the same excited state. The control field and the signal field contribute to a two-photon process coupling two
ground states. In this case the signal pulse is converted into an atomic coherence between ground and excited states and then into a ground state
coherence by the control pulse. However, since the interaction times are very short, it does not allow for the buildup of EIT. This method has been
proposed \cite{Nunn2007,Gorshkov2007} and demonstrated experimentally in far off-resonance conditions \cite{Reim2010,Walmsley2011}.

In this paper, we present a detailed theoretical model for an ultrafast memory without adiabatic approximation and valid for arbitrary frequency
detuning. We show that this protocol holds the promise for a quantum memory with high efficiency, fast operation and broad bandwidth together with
spatial multimode capacity. The problem of the achievable efficiency in this case was treated in Refs.~\cite{Nunn2007,Gorshkov2007} using a numerical
optimization procedure based on the search of the optimal pulse shape for the signal or driving field in the limit of adiabatic elimination of the
excited state. In Ref.~\cite{Gorshkov2008} the shaping of the driving pulse is based on the analysis of the Lagrange function, avoiding the adiabatic
approximation in the optimization procedure. A very good efficiency can be obtained even for short pulse durations. In Ref.~\cite{Golubeva2011} a
different optimization technique based on the minimization of the losses has been used in the resonant case without adiabatic approximation, in the
limit of very short pulses (shorter than the excited state decay time). In the present work we extend this technique to the case of arbitrary
frequency detuning. We show that the memory efficiency can be very good even for large detunings as long as the experimental parameters are properly
optimized. Moreover, we explore the transition region between resonant and adiabatic regimes and we demonstrate that our technique allows finding
optimal parameters for storage.

The article is organized as follows. In Section 2 we present the model system and we write the main equations ruling it. In Section 3 we give the
method for solving the equations for the writing and read-out processes in the semi-classical limit. In Section 4, we study the evolution of the
atomic coherence and of the signal field during the writing process and we calculate the losses. In Section 5, we study the read-out process and the
efficiency of the memory as a whole.

\section{Model system\L{II}}

\begin{figure}
 \centering
 \includegraphics[height=4cm]{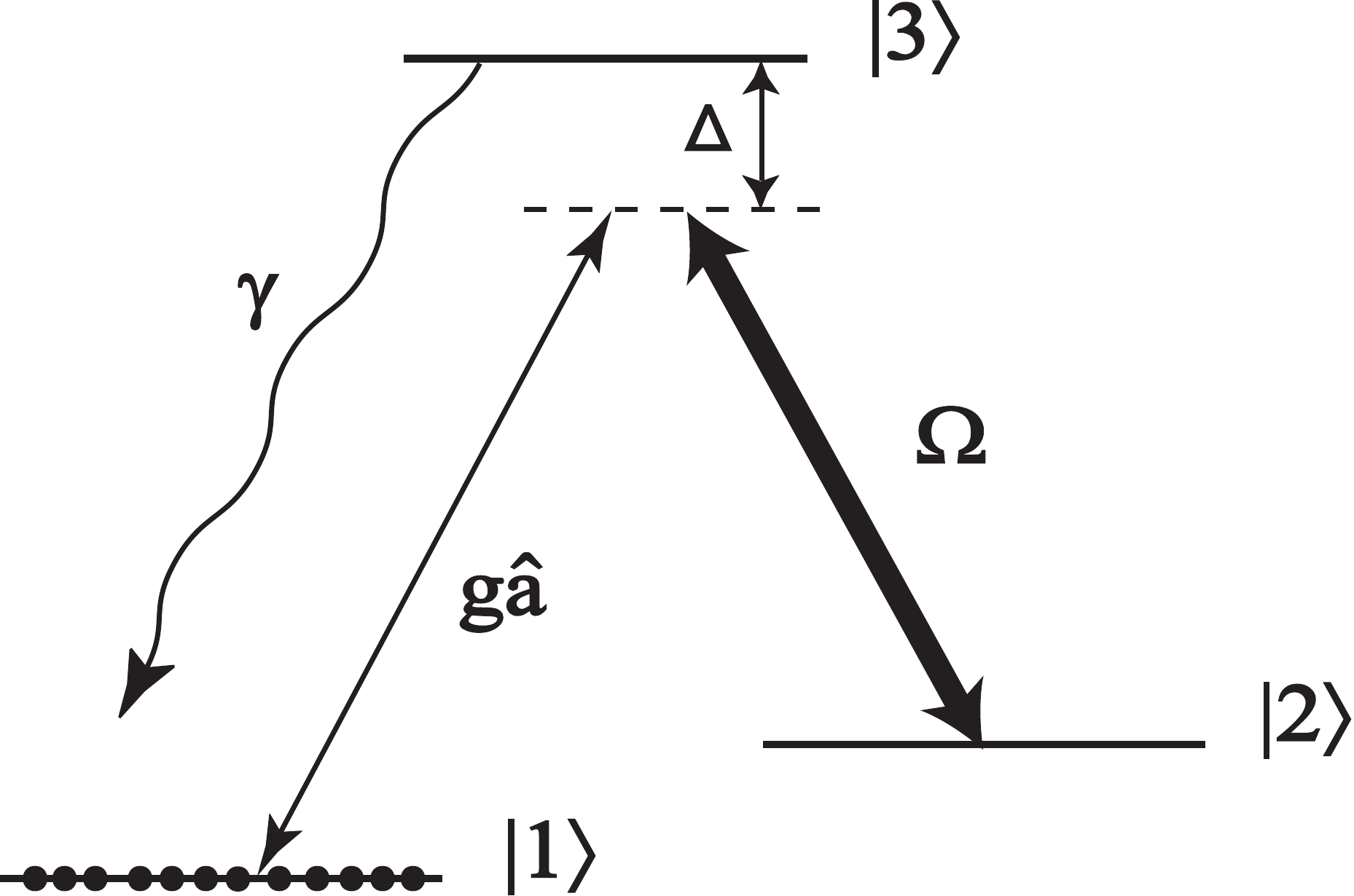}
 \caption{Three level atomic system interacting with driving field $\Omega$ and signal field $a$.}
 \label{Scheme}        
\end{figure}

In this paper, we consider an ensemble of three-level atoms in a $\Lambda$-configuration (Fig.~\ref{Scheme}) that will be used to store temporal and
spatial multimode quantum fields.
The atoms interact with two electromagnetic fields, a signal field
$E_s$ and a driving field  $E_d$, that connect the two atomic ground
states to the excited state. The driving field is a strong,
classical field propagating as plane wave, while the signal field is
a weak quantum field with a transverse structure.

The signal and driving field are very short pulses that are assumed
to be much shorter than the excited state lifetime $\gamma^{-1}$, so
that we can neglect the spontaneous emission during the writing
process.

 In the dipole approximation the light matter
interaction Hamiltonian is given by

\BY
&&\hat V=-\sum_j \hat d_j(t)\hat E(t,\vec r_j), \nn\\
&&\hat E(t,\vec r_j)=\hat E_s(t,\vec r_j)+\hat E_d(t,\vec r_j).\L{2.1}
\EY
Here $\hat d_j(t)$ is the  electric dipole operator of the j-th atom located at $\vec r_j$ . In the paraxial and
quasi-resonant approximations the Hamiltonian can be rewritten in the form
\BY
\hat V=&&\iiint dz \;d^2\rho \[ i\hbar g \[\hat a(z,\vec\rho,t)\hat\sigma_{31} (z,\vec\rho,t)\;e^{\ds i k_s z-i\Delta t}\right.\right.\nn\\
&&\left.\left.-\hat a^\dag(z,\vec\rho,t)\hat\sigma_{13}(z,\vec\rho,t)\;e^{\ds -i k_s z+i\Delta t}\]\right.\nn\\
&& \left.+i\hbar\[\Omega (t)\hat\sigma_{32}(z,\vec\rho,t)\;e^{\ds i k_d z-i\Delta t}\right.\right.\nn\\
&&\left.\left.- \Omega^\ast(t)\hat\sigma_{23}(z,\vec\rho,t)\;e^{\ds - i k_d z+i\Delta t}\]\] .\L{2.2} \EY
Here $k_s$ and $k_d$ are the wave vectors of the signal and driving fields, and $\vec\rho= \vec\rho(x,y)$. The
one-photon detunings of the signal and driving fields are assumed to be equal, and equal to $\Delta $ so that the
two-photon resonance between levels  1 and 2 is fulfilled

 \BY
&&\Delta=\omega_s-\omega_{13}=\omega_d-\omega_{23}.\L{2.3}
\EY
The spatial coordinates $z$ and $\vec\rho$ describe the longitudinal
and  transverse propagation of the signal field. The normalized
amplitude of the signal field $\hat a(z,\vec\rho,t)$ is written as
\begin{eqnarray}
\hat E_s(\vec r,t)&=&-i\sqrt{\frac{\hbar\omega_s}{2\varepsilon_0 c}}e^{-i\omega_s t+ik_sz}\hat a(z,\vec\rho,t) + h.c. ,\L{2.4}
\end{eqnarray}
where $\hat a(z,\vec\rho,t)$ is the annihilation operator for the signal field. Under propagation in free space we have the following commutation
relations \cite{Kolobov1999}
\BY
&&\[\hat a(z,\vec\rho,t),\hat
a^\dag(z,\vec\rho^\prime,t^\prime)\]=\delta^2(\vec\rho-\vec\rho^\prime)\delta(t-t^\prime),\L{2.5}\\
&&\[\hat a(z,\vec\rho,t),\hat a^\dag(z^\prime,\vec\rho^\prime,t)\]=\nn\\
&&\qquad\qquad =c\(1-\frac{i}{k_s}\frac{\partial}{\partial z}-\frac{c}{2k_s^2}\Delta_\perp\)\delta^3(\vec r-\vec r^\prime).\L{2.6}
\EY
The  amplitude $\hat a(z,\vec\rho,t)$ is normalized so that the mean value $\langle\hat a^\dag(z,\vec\rho,t)\hat a(z,\vec\rho,t)\rangle$ is the
photon number per second per unit area. The symbol $\Delta_\perp$ represents the transverse Laplacian with respect to $\vec\rho$
\BY
&& \Delta_\perp=\frac{\partial^2}{\partial x^2}+\frac{\partial^2}{\partial y^2}.\L{2.7}
\EY
We define the intensity of the driving field through the Rabi
frequency $\Omega$. For the sake of simplicity we consider this
value as real $\Omega=\Omega^\ast$. The driving field is a classical
plane monochromatic wave propagating along the z-axis. The coupling
constant between atom and signal field is
\BY
 && g=\(\frac{ \omega_s}{2\epsilon_0\hbar c}\)^{1/2} d_{31}.\L{2.8}
 \EY
where $d_{31}$ is the electric dipole element on the transition
$|1\rangle\to|3\rangle$. We define the collective coherences and
population as sums of over all the atoms
\BY
&&\hat\sigma_{ik}(\vec r,t)=\sum_j\hat\sigma^j_{ik}(t)\;\delta^3(\vec r-\vec r_j)\L{2.9}, \\
&&\hat N_i(\vec r,t)=\sum_j\hat\sigma^j_{ii}(t)\;\delta^3(\vec r-\vec r_j).\L{2.11}
\EY
These quantities fulfill the  commutation relations
\BY
&&\[\hat\sigma_{ik}(\vec r,t),\hat\sigma_{ik}(\vec r^\prime,t)\]=\[\hat N_i(\vec r,t)-\hat N_k(\vec r,t)\]\delta^3(\vec r-\vec r^\prime),\nn\\
&& \L{2.10}
\EY

In this basis one can derive a full system of Heisenberg equations
for the collective operators namely the field amplitude $\hat a(\vec
r,t)$, the collective atomic coherences $\hat \sigma_{ik}(\vec r,t)$
and the collective atomic populations $\hat N_i(\vec r,t)$. Taking
into account the Hamiltonian (\ref{2.2}) and the commutation
relations (\ref{2.6}), (\ref{2.10}), the complete system of the
equations reads
\BY
&& \( \frac{\partial}{\partial t}+c\frac{\partial}{\partial z}-\frac{ic}{2 k_s}\Delta_\perp \)
\hat a=-c g \hat\sigma_{13},\L{12}\\
&& \frac{\partial}{\partial t}{\hat \sigma}_{13}= -i\Delta{\hat \sigma}_{13}+\Omega \hat\sigma_{12} +
 g \hat a (\hat N_1-\hat N_3),\L{13}\\
&& \frac{\partial}{\partial t}{\hat \sigma}_{12}= - \Omega \hat\sigma_{13} - g \hat a
\hat\sigma_{32},\L{14}\\
&& \frac{\partial}{\partial t}{\hat \sigma}_{32}= i\Delta{\hat \sigma}_{32}- \Omega (\hat N_3-\hat N_2) + g \hat a^\dag
\hat\sigma_{12},\L{15}\\
&&\frac{\partial}{\partial t}{\hat N}_1= - g \hat a \hat\sigma_{31} -
g \hat a^\dag  \hat\sigma_{13},\L{16}\\
&& \frac{\partial}{\partial t}{\hat N}_2= - \Omega \(\hat\sigma_{32} -
 \hat\sigma_{23}\),\L{17}\\
&& \frac{\partial}{\partial t}{\hat N}_3=-\frac{\partial}{\partial t}{\hat N}_1-\frac{\partial}{\partial t}{\hat N}_2.\L{18}
\EY
To derive the equations (\ref{12})-(\ref{18}) we have performed the substitutions
\BY &&\hat \sigma_{13} \to e^{\ds i k_s z-i\Delta t} \hat \sigma_{13},\nn\\
&&\hat \sigma_{23} \to e^{\ds i k_d z-i\Delta t} \hat \sigma_{23},\L{2.18}\\
&&\hat \sigma_{12} \to e^{\ds -i (k_d-k_s) z} \hat \sigma_{12}.\nn
 \EY

We have omitted the terms related to spontaneous relaxation $|3\rangle\to|1\rangle$ since we have assumed that the relaxation rate $\gamma$ is small
enough for the spontaneous emission to be negligible during the short time duration of the pulses. The
 equations can be written in a simplified way with a few
 approximations given below.

According to equations (\ref{2.11}) the collective atomic operators
have sharp spatial distributions due to the delta-localization of
the atoms. However, due to the collective effect of atoms localized
along a field trajectory, we can average Eqs.~(\ref{12})-(\ref{18})
over the positions of the atoms.

We will also replace the operator $\overline{\hat N}_1-\overline{\hat N}_3 $ in Eq.~(\ref{13}) by the number giving the mean atomic density $N$. We
will assume that in the beginning of the process most atoms are in state $|1\rangle$. During the memory processes (writing and read-out) the
population of the state $|1\rangle$ stays close to its initial value, because the photon number in the signal pulse is much smaller than the initial
atomic number.

In Eq.~(\ref{14}), we  can neglect the second term on the right hand
side since $g\hat a$ is much smaller than $\Omega$ (we have assumed
 $|\Omega|^2\gg g^2\langle \hat a^\dag\hat a\rangle$).
Furthermore, $\hat \sigma_{32}\ll\hat \sigma_{13}$ because the
populations $N_2$ and $N_3$ are much smaller than $N$.

Now we can write a simplified system of partial differential
equations describing the evolution of the system as
\BY
&&  \(\frac{1}{c}\frac{\partial}{\partial t}+\frac{\partial}{\partial z}-\frac{i}{2 k_s}\bigtriangledown_\perp^2 \) \hat a(z,\vec\rho,t)= - g\;
\hat\sigma_{13}(z,\vec\rho,t), \quad\L{2.12}\\
&& \frac{\partial}{\partial t}{\hat\sigma}_{13}(z,\vec\rho,t)= -i\Delta\hat\sigma_{13}(z,\vec\rho,t)\nn\\
&& \qquad\qquad\qquad\quad +  gN \hat a(z,\vec\rho,t)+\Omega\hat\sigma_{12}(z,\vec\rho,t),\L{2.13}\\
&& \frac{\partial}{\partial t}{\hat\sigma}_{12}(z,\vec\rho,t)=  - \Omega\hat\sigma_{13}(z,\vec\rho,t).\L{2.14}
\EY
In a previous publication \cite{Golubeva2011}, similar equations
were derived for the case of a resonant excitation ($\Delta =0$).
Here and below we omit the averaging over the atomic localizations.

Let us renormalize the coherences $\hat\sigma_{13}$ and
$\hat\sigma_{12}$
\BY
&&\hat\sigma_{12}(z,\vec\rho,t)/\sqrt N=\hat b(z,\vec\rho,t),\\
&&\hat\sigma_{13}(z,\vec\rho,t)/\sqrt N=\hat c(z,\vec\rho,t)\L{2.15}
\EY
so that they obey to the bosonic commutation relations:
\BY
&&\[\hat b(\vec r,t),\hat b^\dag(\vec r^\prime,t)\]=\[\hat c(\vec r,t),\hat c^\dag(\vec r^\prime,t)\]= \delta^3(\vec r-\vec r^\prime).\qquad\L{2.16}
\EY
Here again we have taken into account the fact that $\hat N_1-\hat
N_{2,3}\to N$.

In the Fourier domain relative to the transverse coordinates
$\vec\rho $ the equations read
\BY
&& \frac{\partial}{\partial z} \hat a(z,t;\vec q)=- g\sqrt N \;\hat c(z,t;\vec q),\L{2.19}\\
 &&\frac{\partial}{\partial t}\hat c(z,t;\vec q)=-i\Delta\hat c(z,t;\vec q)\nn\\
 &&\qquad\qquad\qquad+   g\sqrt N \;\hat a(z,t;\vec q)
 +\Omega \hat b(z,t;\vec q),\quad\;\;\;\L{2.20}\\
 && \frac{\partial}{\partial t}\hat b(z,t;\vec q)= - \Omega\hat c(z,t;\vec q)\L{2.21},
\EY
where we have introduced the transverse wavevector $\vec q$ and we have made the changes
\BY
 && \hat a(z,t;\vec q)\to\hat a(z,t;\vec q)e^{\ds -i q^2 z/(2k_s)},\L{2.22.}\\
 && \hat b(z,t;\vec q)\to
 \hat b(z,t;\vec q)e^{\ds -i q^2 z/(2k_s)},\L{2.22}\\
 && \hat c(z,t;\vec q)\to\hat c(z,t;\vec q)
  e^{\ds -i q^2 z/(2k_s)}.\L{2.22}
 \EY
From the system (\ref{2.19})-(\ref{2.21}) one can obtain a conservation equation
\BY
&&\frac{\partial \hat a^\dag\hat a}{\partial z}+\frac{\partial \hat b^\dag\hat b}{\partial t}+\frac{\partial \hat c^\dag\hat c}{\partial t}=0.\L{30}
\EY
This equation means that the input photons of the  weak quantum field are converted into excitations of the atomic coherences $\hat\sigma_{13}$ and
$\hat\sigma_{12}$. The aim is to store the information carried by the signal field in the ground state coherence $\hat\sigma_{12}$, so that an
excitation of the state $|3\rangle$ is undesirable. It is possible to reduce this loss channel by an appropriate choice of the driving field. If the
driving field power is high enough for the Rabi oscillation on the transition $|2\rangle\to|3\rangle$ to be more effective than a spontaneous
emission ($\Omega\gg\gamma$) and if the pulse duration is short enough, so that the atoms undergoing a Rabi oscillation have no time to go back to
the state $|3\rangle$, then the third term in Eq.~(\ref{30}) should be negligible. This will be studied in the optimization of the memory process.

We neglect the time delay linked to the pulse propagation in the
atomic medium. This means that, if we have long enough pulses, such
that $L/c\ll T$ ($L$ is the thickness of the medium and $T$ is the
pulse duration), we can neglect the time interval between the time
at which the front part of the pulse enters the medium and the time
at which the front part leaves it. Formally this means we can
neglect the time derivative in Eq.~(\ref{2.12}). For simplicity we
will assume that the driving pulse has a rectangular time
distribution (in the equations $\Omega(t)=const$ for $0<t<T$).

\section{Writing and read-out processes in the semi-classical limit\L{IV}}

The main aim of this paper is the determination of the memory
efficiency. As it is well known the semiclassical description is
sufficient for this and we can use the initial conditions :
$b(0,z;\vec q)=c(0,z;\vec q)=0$ for writing and $a_{in}(t;\vec
q)=c(0,z;\vec q)=0$ for read-out. Here and below we omit the
operator notation for the variables. The detailed resolution of the
system of  partial differential equations (\ref{2.19})-(\ref{2.21})
can be found in Apps.~\ref{A},\ref{B}.

Using the general solutions (\ref{A1})-(\ref{A3}) one can obtain the
semi-classical ones for the writing process for both the field
amplitude $a^W(t,z;\vec q)$ and the atomic coherence $b(t, z;\vec
q)$ in the form
 \BY
&& a^W(\tilde t,\tilde z;\vec q)=  \int_0^{\tilde T_W} d\tilde t^\prime a_{in}(\tilde t^\prime,\vec q) G_{aa}(\tilde
t-\tilde t^\prime, \tilde z),\L{3.1}\\
&& b^W (\tilde t, \tilde z;\vec q)=
 -p\int_0^{\tilde T_W} d\tilde t^{\prime}a_{in}(\tilde t^\prime,\vec q)
 G_{ab}(\tilde t-\tilde t^{\prime}, \tilde z ),\L{3.2}
\EY
where we have introduced the dimensionless time $\tilde t$ and
longitudinal spatial coordinate $\tilde z$ according to
\BY
&&\tilde t =\Omega \;t,\qquad\qquad\tilde T_W=\Omega T_W,\nn\\
&&\tilde z =\frac{2g^2N}{\Omega} \;z,\qquad \tilde L =\frac{2g^2N}{\Omega} \;L
\EY
where $T_W,$ is the common duration of signal and driving pulses for
writing and $L$ is the thickness of the memory cell. We have defined
an effective interaction coefficient $p$ given by
\BY
&& p=\frac{g\sqrt N}{\Omega}.
\EY
 The kernels $G_{aa}(t,z)$ and $G_{ab}(t,z)$ are time convolutions
\BY
&&G_{aa}(\tilde t,\tilde z)=\int_0^{\tilde t}d\tilde t^\prime f(\tilde t^\prime,\tilde z;r)f^\ast(\tilde t-\tilde
t^\prime,\tilde z;-r),\\
&&G_{ab}(\tilde t,\tilde z)=\int_0^{\tilde t}d\tilde t^\prime f_0(\tilde t^\prime,\tilde z;r)f_0^\ast(\tilde t-\tilde t^\prime,\tilde z;-r),
\EY
where the functions $f$ and $f_0$ are expressed via the n-th Bessel function of the first kind denoted by $J_n$:
\BY
 f(\tilde t,\tilde z;r)=&&\delta(\tilde t)-e^{\ds -ir\tilde t}e^{\ds -i\sqrt{1+r^2}\;\tilde t}\nn\\
 &&\times\sqrt{\frac{(1+r)\tilde z}{4\tilde t}}\;J_1\(\sqrt{(1+r)\tilde z\tilde t}\)\Theta(\tilde t),\\
 f_0(\tilde t,  \tilde z;r )=&&e^{\ds -ir\tilde t}e^{\ds -i\sqrt{1+r^2}\;\tilde t}\;J_0 \(\sqrt{(1+r)\tilde z\tilde t}\) \Theta(\tilde t).\nn\\
 &&
\EY
Here the frequency detuning is given by the dimensionless parameter $r=\Delta/(2\Omega)$ and $\Theta(\tilde t)$ is the step function $\Theta(\tilde
t)=1$ for $0<\tilde t<\tilde T_W$ and equals zero
 otherwise.

In order to estimate the efficiency of the memory, we calculate the
field amplitude at the output of the cell. This amplitude is
different in the case of forward and backward retrieval. The
corresponding expressions read
\BY
&&a^R_{for}(\tilde t,\tilde L;\vec q)= \\
&&\qquad=\frac{1}{2}\int_0^{\tilde T_W} d\tilde t^{\prime}a_{in}(\tilde t^\prime,\vec q)\int_0^{\tilde L} d\tilde z G_{ab}(\tilde t, \tilde
z)G_{ba}(\tilde t^{\prime}, \tilde L-\tilde z ),\nn\\
&&G_{ba}(\tilde t, \tilde z)=G_{ab}(\tilde t, \tilde z),\nn
\EY
and
\BY
&&  a^R_{back}(\tilde t,\tilde L;\vec q)= \\
&&\qquad=\frac{1}{2}\int_0^{\tilde T_W} d\tilde t^{\prime}a_{in}(\tilde t^\prime,\vec q)\int_0^{\tilde L} d\tilde z G_{ab}(\tilde t, \tilde
z)G_{ba}(\tilde t^{\prime}, \tilde z ).\nn
\EY
The last two formulas are correct only in the approximation where diffraction is neglected. Indeed, we have not taken
into consideration the diffraction phenomenon described by equation (\ref{2.22.})-(\ref{2.22}). As was demonstrated
in \cite{Golubeva2011} this effect does not introduce any significant corrections in the case of the forward
retrieval but restricts  the mode number for the backward retrieval.

In the following we will use an optimization procedure based on the
choice of the optimal relation between the thickness of the memory
cell $\tilde L$ and the duration $\tilde T_W$ of the signal and
driving pulses. We will compare this optimization approach with the
approach developed in Ref.~\cite{Gorshkov2007} based on the
optimization of the driving pulse shape in the case of short pulses.

\section{Discussion of the writing process \L{V}}

\begin{figure*}[h]
 \centering
\includegraphics[height=50mm]{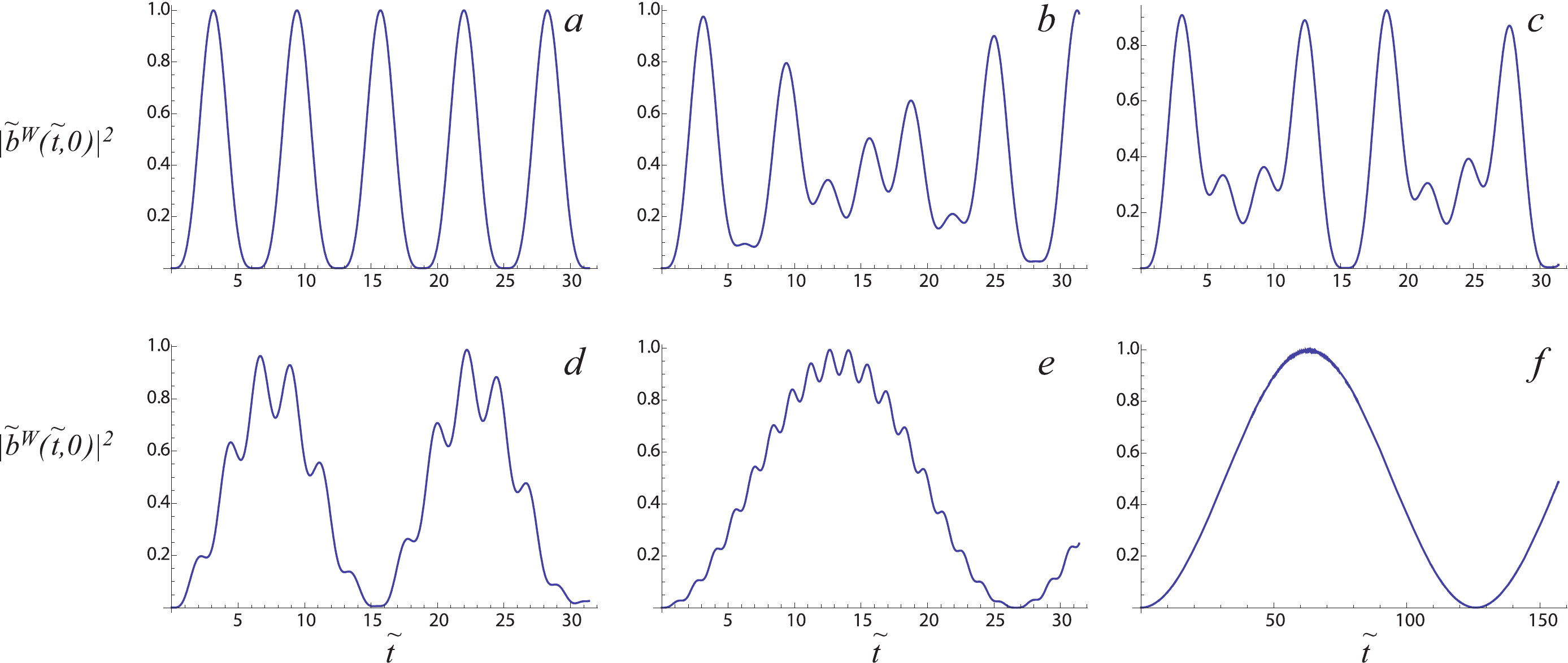}
   \caption{Normalized coherence at $\tilde z=0$ as a function of time
 for (a) $r=0$,(b) $r=0.1$, (c) $r=0.2$, (d) $r=1$, (e) $r=2$, (f) $r=10$. }
  \label{fig:z=0}
\end{figure*}

In this section, we study the writing process, i.e. the conversion
of the signal field into atomic coherence.  Let us start with a
simple calculation of this process at the input of the memory cell.
For this we solve Eq.~(\ref{3.2}) at $\tilde z=0$. In this case the
 atomic polarization $ b^W (\tilde t,0;\vec q)$ corresponding to the "written" information reads
\BY
&&\tilde b^W (\tilde t,0)=b^W (\tilde t,0;\vec q)/(-2p a_{in}(\vec q))=\nn\\
&&\qquad\quad\;\;\; =\frac{1}{2}\(1-e^{-i r \tilde t}\[\cos (\tilde t\sqrt{r^2+1}) \right.\right.\nn\\
&&\qquad\qquad\;\;\;\; +\left.\left.\frac{i r}{\sqrt{r^2+1}}\sin (\tilde t\sqrt{r^2+1})\]\).\L{z=0}
\EY
For the sake of the simplicity we have taken $a_{in}(\tilde t,\vec
q)=const (\tilde t)$. The normalized atomic polarization ranges from
0 to 1. In the limit of small or large detuning, we have a simple
behaviour
\BY
&&r\ll1:\qquad|\tilde b^W (\tilde t,0)|^2=\sin^4\frac{\tilde t}{2},\\
&&r\gg1:\qquad|\tilde b^W (\tilde t,0)|^2=\sin^2\frac{\tilde t}{4r}
\EY
Fig.~\ref{fig:z=0}a shows the Rabi oscillation for a normalized
detuning $r=0$ while Fig.~\ref{fig:z=0}f shows a periodical
modulation with a period of $4\pi r$ for $r=10$. When $r$ increases,
one can see in Fig.\ref{fig:z=0}b,c the beat between two close
frequencies. For larger $r$ (Fig.\ref{fig:z=0}d,e) the curves show a
slow oscillation (that corresponds to the term with frequency
$\sqrt{r^2+1}-r$ in (\ref{z=0})), modulated by a fast oscillation
(at frequency $\sqrt{r^2+1}+r$), with a modulation depth decreasing
with increasing $r$.

\begin{figure*}
 \centering
\includegraphics[height=35mm]{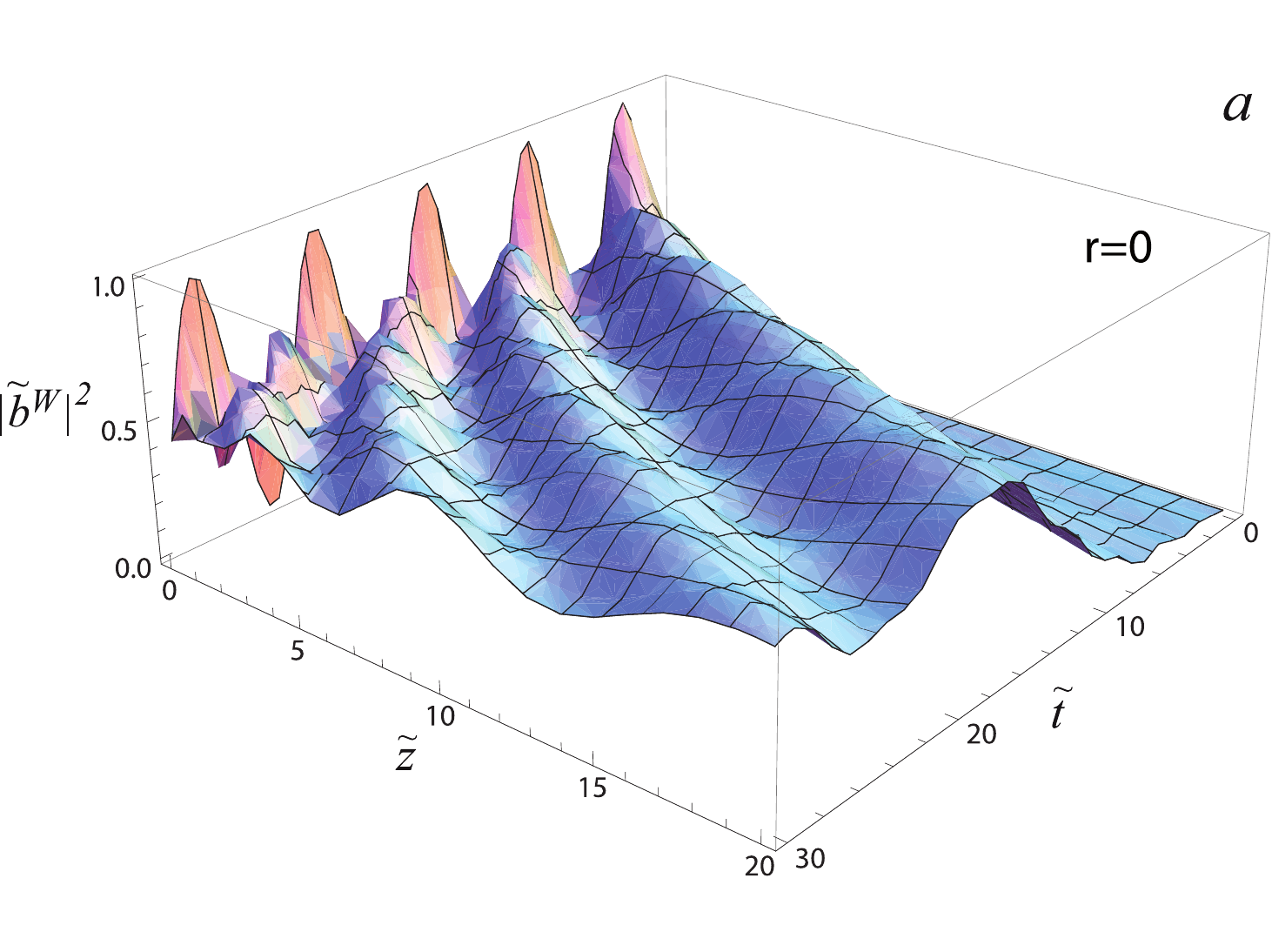}\qquad\qquad
\includegraphics[height=35mm]{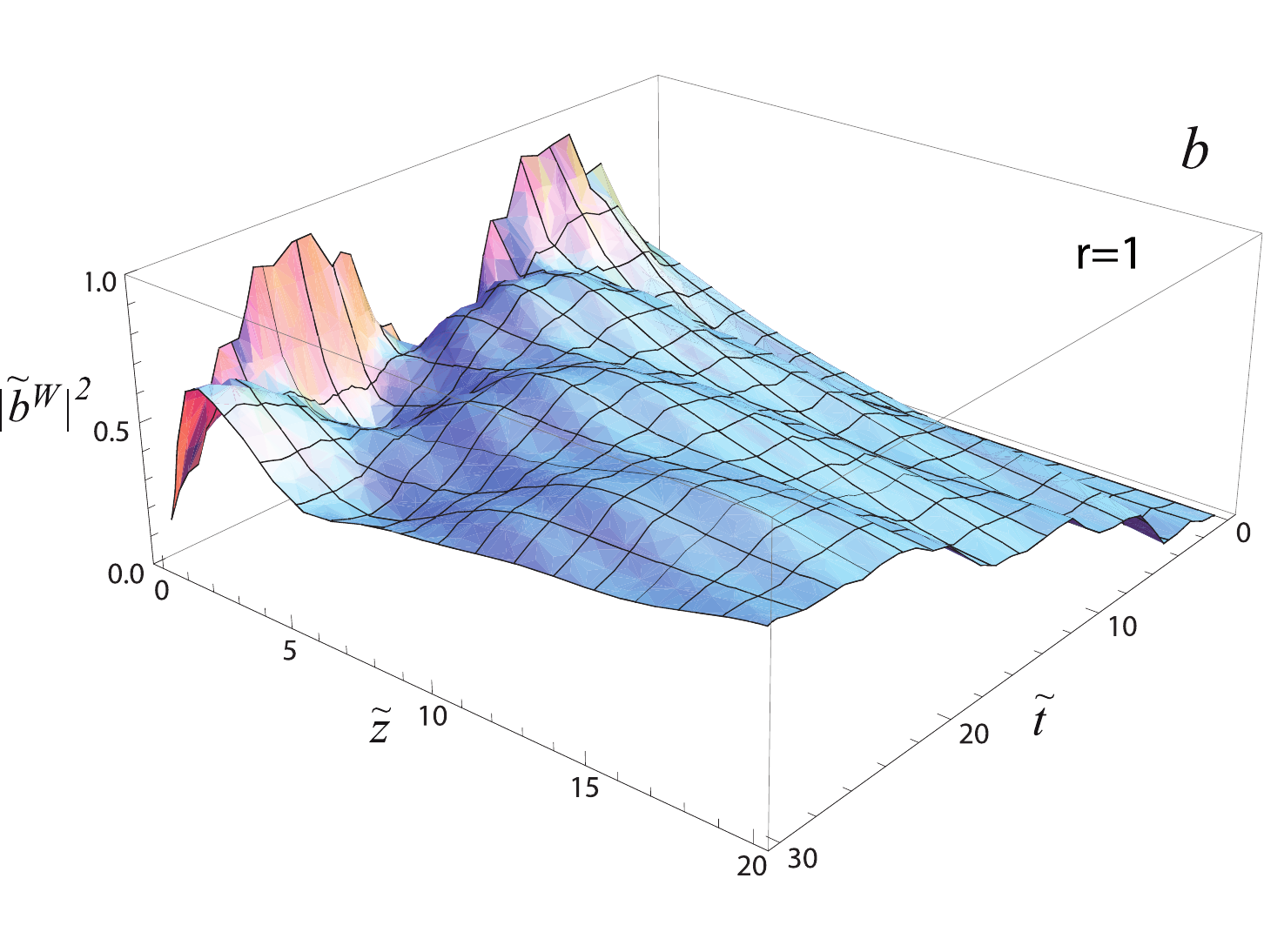}\qquad\qquad
\includegraphics[height=35mm]{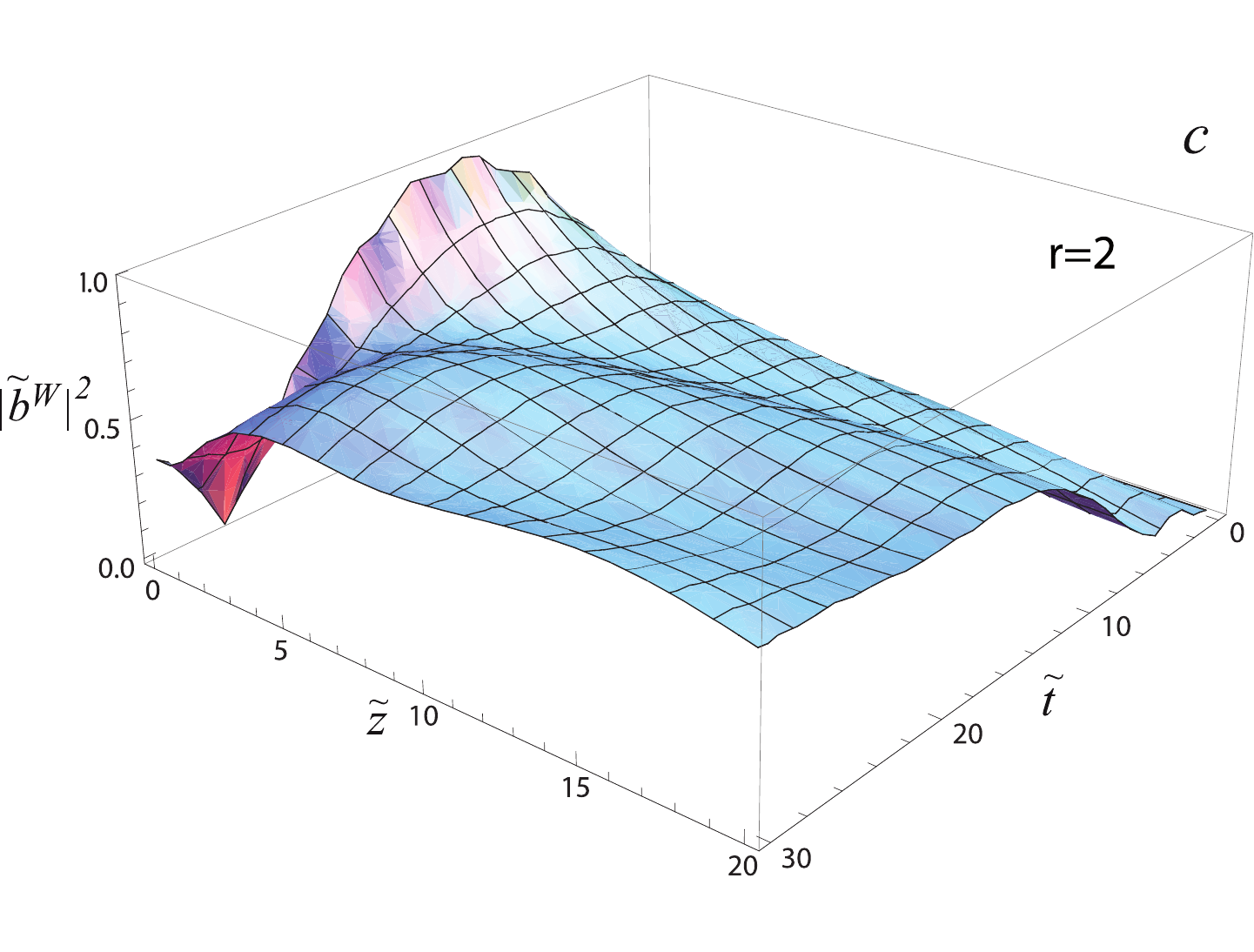}
 \caption{Distribution of the coherence $|\tilde b^W(\tilde t, \tilde z)|^2$ in time and space for (a)$r=0$, (b) $r=1$ and (c) $r=2$. }
  \label{fig:3D}
\end{figure*}

Following the variation of $|\tilde b^W(\tilde t,0)|^2$ as a
function of $r$ allows to get a first view of the behaviour of the
system when the detuning is varied. It can be seen from
Fig.~\ref{fig:z=0} that even for a small detuning, $r=0.1$, a
significant distortion of the coherence profile at the input of the
medium takes place as compared to the resonant case. On the other
hand for $r=2$ the high frequency modulation of the slow
oscillations is rather small and the excitation can be considered as
close to the off-resonant case, where the well-known solutions in
the Raman limit can be used. A detailed study of the coherence
distribution for all $\tilde z$, as given below, will allow to
better characterize the interaction regime, between resonant and
Raman.

\begin{figure*}[h]
 \centering
  \includegraphics[height=40mm]{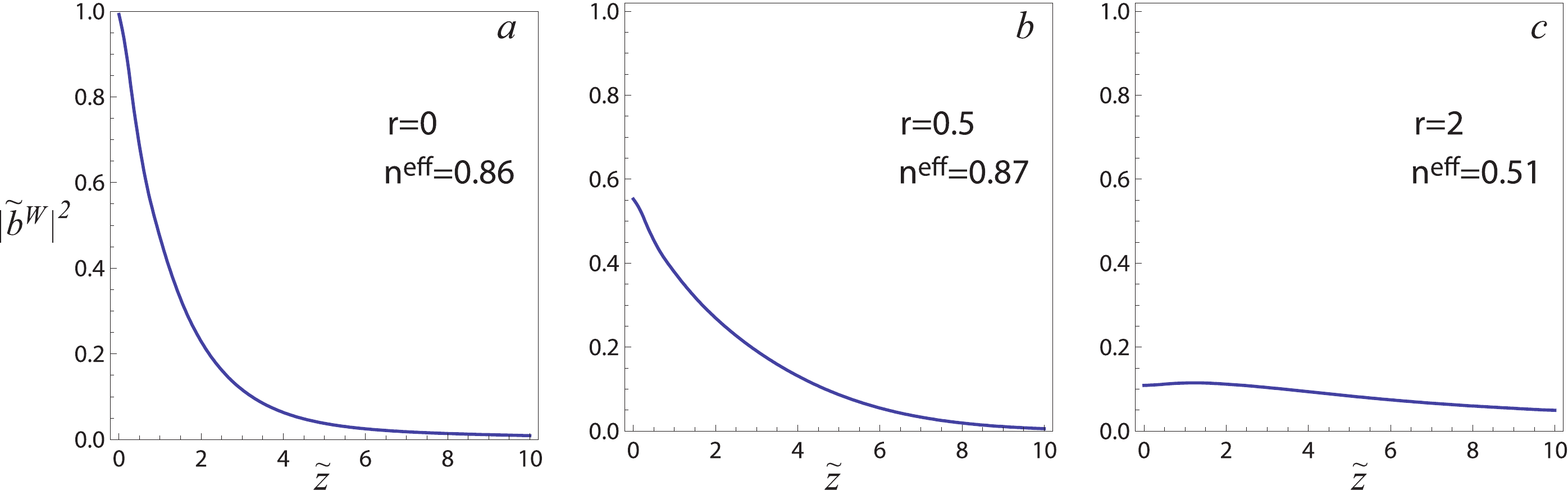}
   \caption{Normalized distributions of the atomic coherence inside the medium
 at time $\tilde t=\pi$ for (a)$r=0$, (b) $r=0.5$, (c)$r=2$. }
  \label{fig:coherence}
\end{figure*}

When shifting into the medium, the behavior of the coherence is much
more complicated since it looses its simple harmonic character.
Figure \ref{fig:3D} shows a displacement of the maximum of the
coherence along the medium. Let us follow the dependence of $|\tilde
b^W (\tilde t,\tilde z)|^2$ on $\tilde z$ for a given value of
$\tilde t=\tilde T^W=\pi$ given in Fig. \ref{fig:coherence}.

Comparing the curves for different values of the detuning, we see
that the value of coherence at $\tilde z=0$ significantly decreases
with increasing detuning. From these curves it could be concluded
that the larger $r$ the less information is written in the cell.
However, we will show that it is not true. Let us introduce the
quantity $n^{eff}$ that characterizes the proportion of signal
photons converted into coherence $\tilde b^W$ during the writing
process:

\BE
n^{eff}(\tilde T^W,\tilde L)=\frac{1}{\tilde T^W}\frac{1}{2}\int_0^{\tilde L} |\tilde b^W (\tilde T^W,\tilde z)|^2 d\tilde z.\L{n_eff}
\EE
Here, $1/\tilde T^W$ before the integral comes from the input pulse
energy, and the factor $1/2$ comes from the previously introduced
dimensionless variables. The integral gives the normalized
population $N_2$ in the medium with length $\tilde L$ during the
interaction time $\tilde T^W$. Since the transition of an atom to
the level $|2>$ in our model corresponds to the coherent scattering
of a photon from the signal wave, this is also the number of signal
photons that was recorded in the atomic coherence. We are interested
in the percentage of input photons recorded in such a way. The
calculated values of $n^{eff}$ are presented Fig.
\ref{fig:coherence}a-c. We see that the number of recorded photons
in the first two panels are almost identical (there is even a small
increase of $n^{eff}$ for $r=0.5$ as compared with the resonant
value), while for $r=2$ the value $n^{eff}$ decreases by a factor of
less than 2.

\begin{figure*}[h]
 \centering
  \includegraphics[height=40mm]{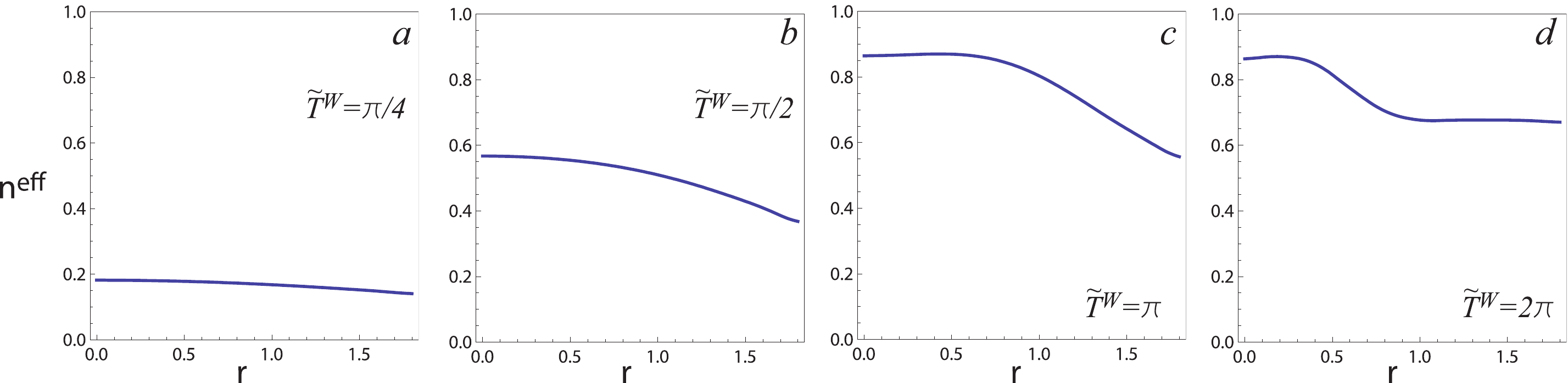}
 \caption{Fraction of signal photons (normalized to the energy of input signal pulse) that have been converted to atomic coherence $b^W$ during
writing as a function of the detuning parameter $r$ for $\tilde
L=10$ and (a) $\tilde T^W=\pi/4$, (b) $\tilde T^W=\pi/2$, (c)
$\tilde T^W=\pi$, (d) $\tilde T^W=2\pi$.}
  \label{fig:n_eff}
\end{figure*}
We can also follow the dependence of $n^{eff}$ on $r$ for a given
medium length and for different durations of the signal pulse. Fig.
\ref{fig:n_eff} shows that the proportion of recorded photons
decreases when the detuning increases, but this decrease depends on
the duration of the writing process. In particular, we see that the
writing efficiency depends weakly on the detuning for $\tilde
T^W=\pi/4 $ but it stays quite low. With increasing pulse duration
($\tilde T^W=\pi$) a plateau appears for detuning range 0 to 0.7.
This range is reduced for $\tilde T^W=2\pi$, but a second
 plateau appears for $r\in[1, 1.8]$.
 While Figs. \ref{fig:3D}, \ref{fig:coherence} and \ref{fig:n_eff}
 give a detailed behaviour of the efficiency of the writing process
 depending on the detuning, on the time duration of the pulse and on
 the length of the medium, it appears that the optimization of the
 efficiency is non trivial and requires a specific procedure. This
 will be studied in the next section.

\subsection{Estimation of the writing losses  \L{VI}}

In Ref.\cite{Golubeva2011} an algorithm of memory optimization,
based on the minimization of leakage is described. Leakage is
defined as

\begin{figure*}[h]
 \centering
\includegraphics[height=40mm]{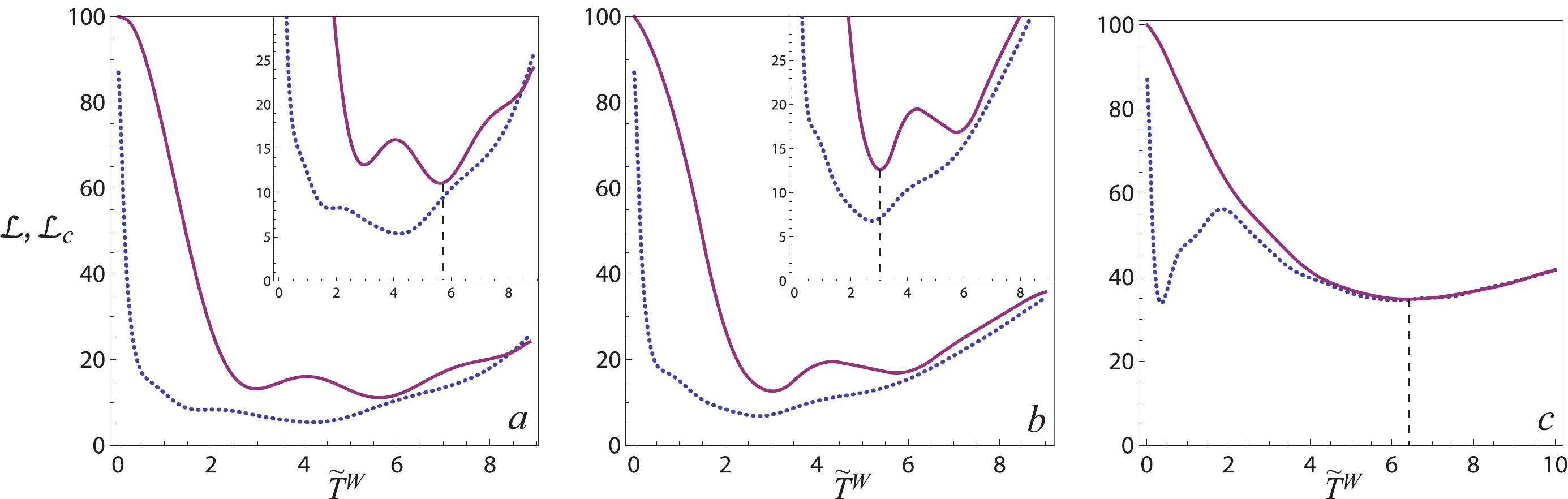}
   \caption{Writing process : relative losses associated with leakage (blue curves, dotted lines) and relative total losses (red curves, full lines)
    (in percent of the input field  intensity) at the output
 of the medium as a function of $\tilde T_W$ for
  $\tilde L=10$ for (a)$r=0$, (b) $r=0.5$, (c)$r=2$. }
  \label{fig:losses}
\end{figure*}

\BE
{\cal L}(\tilde T^W,\tilde L)=\frac{\int_0^{\tilde T_W}|a^W(\tilde t,\tilde L)|^2 d \tilde t}{\int_0^{\tilde T_W}| a_{in}(\tilde t)|^2  d \tilde
t}\times100\%,\L{Loss}
\EE
It characterizes the proportion of signal photons going out of the
cell during the writing time. Such  an estimation of the losses is
justified when the leakage is the main origin of losses. However, as
will be shown below, there is a range of $\tilde T^W$ in which the
population of the upper energy level is large enough to cause
significant losses. We will show that for a high-speed memory
 the three-level atomic system can not be reduced
to a two-level scheme. Note that this situation is specific of the
case of simultaneous interaction of the signal and control fields
with matter, and does not happen in memory protocols based on an
echo \cite{Moiseev2001}. We have already introduced the value
$n^{eff}$, which is the proportion of signal photons that have been
recorded. Then, the value of the total losses (as a percentage of
the number of photons in the input signal pulse) can be expressed as
follows :

\BE
{\cal L}_c(\tilde T^W,\tilde L)=(1-n^{eff}(\tilde T^W,\tilde L))\cdot 100\%.\L{comp_los}
\EE
Fig. \ref{fig:losses} shows the losses associated with leakage ${\cal L}$  (blue curves, dotted lines) and the total losses of photons ${\cal L}_c$,
(red curves, full lines) as a function of the duration of the writing for a given medium length, and three different values of $r$.

First, let us note that these curves are not monotonous and exhibit
one or several minima. This means that for a given length $\tilde L$
one can find the pulse duration which is recorded in the atomic
medium with minimal losses. Significant difference between the
curves corresponding to leakage and to total losses come from the
role of the upper level in the interaction of such pulses with the
atomic medium. However, in the region of minimum losses, the
distance between the two curves is small. The optimization of the
memory based on leakage allows to define a range of values $\tilde
T^W$, for which an efficient writing is expected. However, the curve
giving the total losses is a more precise tool to determine the
optimum ratio between $\tilde T^W$ and $\tilde L$. As can be seen
from the plots, above some value of $\tilde T^W$ the two curves
coincide, i.e. all the system losses are associated  with leakage
only and level of $|3\rangle$ is not populated at the end of the
process. The larger detuning $r$ the smaller the value $\tilde T^W$
for which this happens.

\begin{figure*}[h]
 \centering
 \includegraphics[height=80mm]{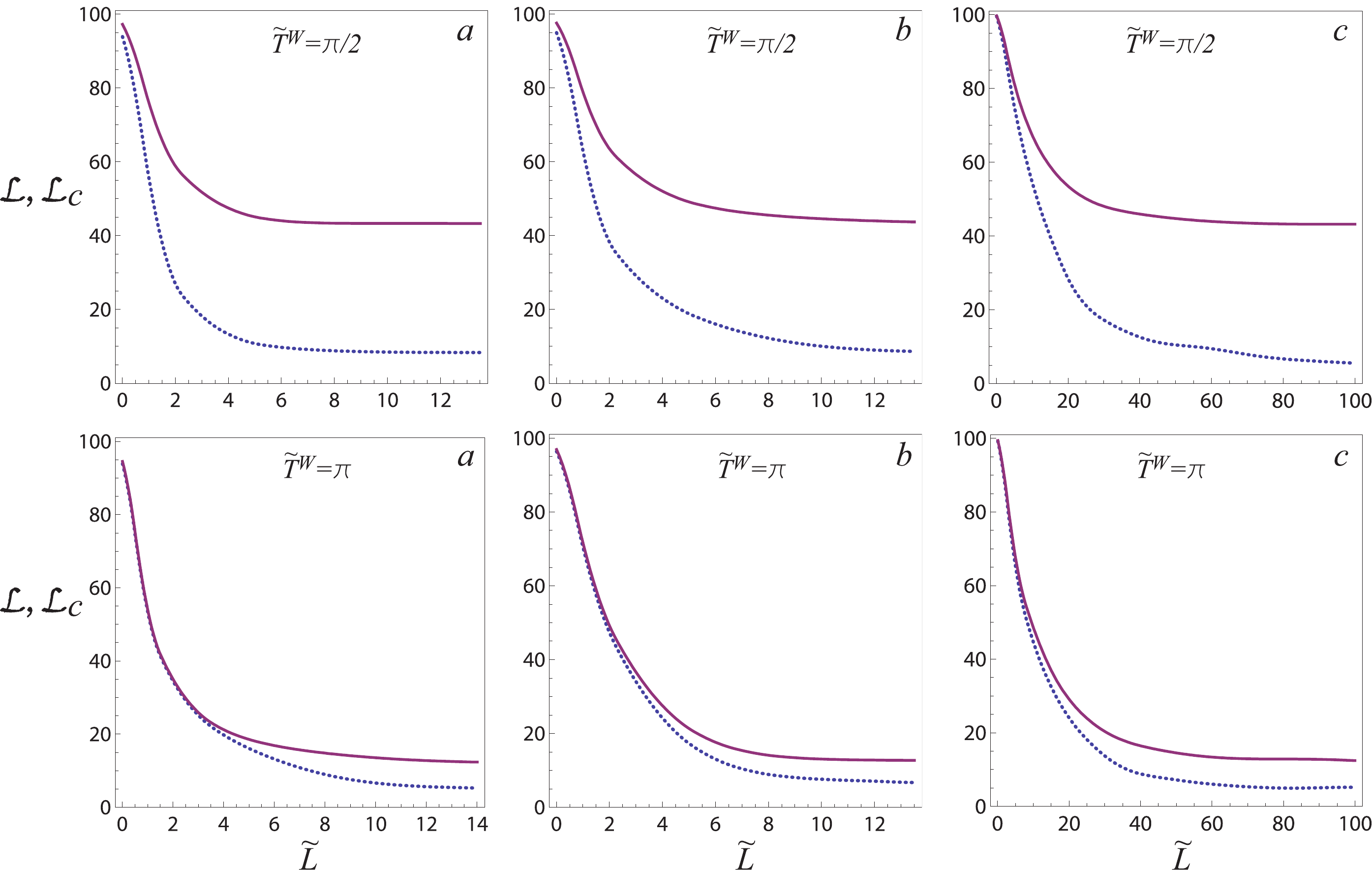}
   \caption{Writing process : relative losses due to leakage (blue curves, dotted lines) and relative total losses (red curves, full lines)
   (in percent of the input field intensity) at the output
 of the medium as a function of $\tilde L$ and (a) $r=0$, (b) $r=0.5$ and (c) $r=2$ for $\tilde T_W=\pi/2$ (first row) and
 $\tilde T_W=\pi$ (second row). }
  \label{fig:lossesL}
\end{figure*}

We can also follow the dependence of the losses on the length $\tilde L$ for a given value of $\tilde T^W$ as shown in Fig. \ref{fig:lossesL}.
These curves have a monotonous variation, and show that the
efficiency increases with increasing medium length. When the pulse
duration increases, the curves of ${\cal L}$ and ${\cal L}_c$ get
closer to each other, in agreement with the analysis of Fig.
\ref{fig:losses}. A specific feature of these curves is their
saturation for large values of $\tilde L$. The presence of a plateau
on the leakage plots is actually due to an approximation made in the
model; we have neglected the time intervals associated with the
propagation of the pulse wavefronts inside the medium, and we
assumed $\tilde t=0$ is the time at which the wavefront reaches the
the cell output, while $\tilde t=\tilde T^W$ is the time at which
the end part of the pulse arrives at the entrance of the cell. This
means that we always have some leakage in the initial time
independently of the length of the medium.

The difference between the levels of the plateau for ${\cal L}$ and ${\cal L}_c$ characterizes the losses due to the non zero population of level
$|3\rangle$. One can see that this value is constant for large enough length $\tilde L$. This value saturates because of the depletion of the signal
field, so that further increase of the medium length cannot change the populations $N_2$ and $N_3$. The difference between ${\cal L}$ and ${\cal
L}_c$ depends strongly on the pulse duration. In particular if the pulse is too short, many atoms are left in the upper state. It can also be seen in
Figs. \ref{fig:lossesL}c that when the detuning increases, the saturation occurs at larger values $\tilde L$, and that a high efficiency can be
reached as well if the atomic medium is long enough.
Various optimizations procedures have been proposed in the limit when the excited state can be adiabatically eliminated \cite{Nunn2007,Gorshkov2007}.
In particular, the optimization used in reference \cite{Gorshkov2007} allows to get the maximum available efficiency for long enough durations of the
writing process, but breaks down when the duration of a writing process $T^W$ gets smaller than the excited state decay time $\gamma^{-1}$ divided by
the optical depth of the medium $d$. In Ref. \cite{Gorshkov2008} the numerical optimization procedure relying on the shaping of the driving pulse was
extended to the non adiabatic case, which allowed to reach better storage efficiency for short pulses. The latter technique was developed for the
resonant case, yielding optimal memory efficiencies that are very close to the ones presented here. The applicability of our optimization method to
various detunings shows that such a memory can be also very efficient in the off-resonant regime, bringing more flexibility for experimental
realizations.

\subsection{Validity limits of resonant and Raman approximations  \L{VII}}

The solutions that we have presented are valid for detunings ranging
from zero to large values that correspond to the case of Raman
interaction, where the system is effectively reduced to a two-level
system. General formulas covering the full range of detunings allow
a comparison with the limit cases of resonant and Raman
interactions. We can identify the largest detuning for which the
resonant approximation is still valid, yielding the same storage
efficiency. On the other hand, we can estimate for which value of
$\Delta$ a simplified Raman description can be used without yielding
appreciable errors in the memory efficiency.

\begin{figure*}[h]
 \centering
\includegraphics[height=40mm]{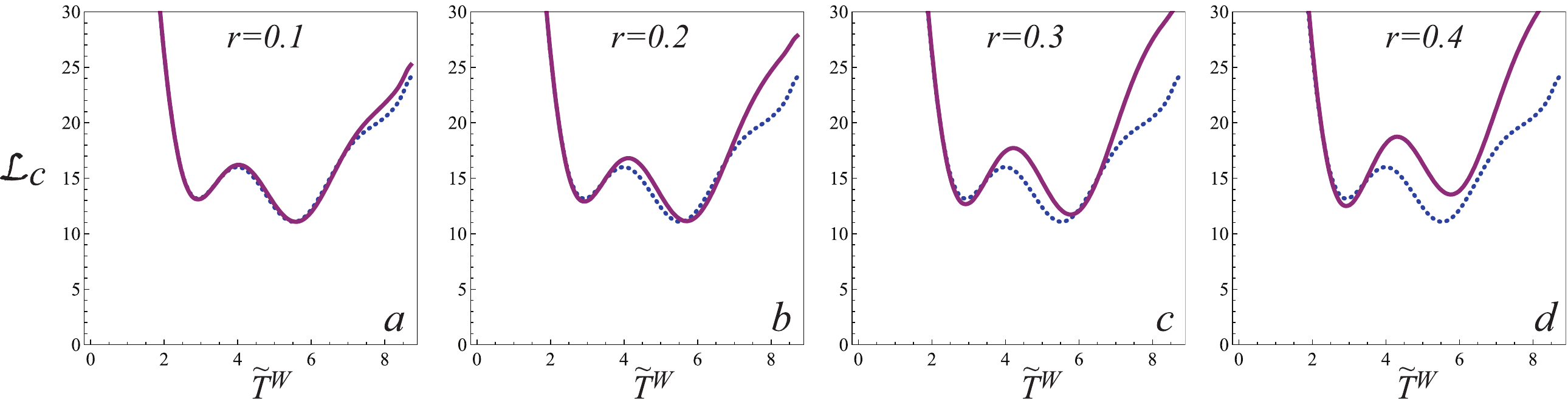}
   \caption{Comparison of relative total losses (in percent of the input field intensity) at the output
 of the medium as a function of $\tilde T_W$ for resonant (blue curves, dotted lines) and detuned (red curves, full lines) cases;
  $\tilde L=10$, (a)$r=0.1$, (b) $r=0.2$, (c)$r=0.3$, (d)$r=0.4$.}
  \label{fig:r=0.1}
\end{figure*}

In Fig. \ref{fig:z=0}b one can see a significant distortion of the
temporal profile of the atomic ground state coherence at the input
of the medium for $r=0.1$ as compared to the resonant case. However
this detuning does not affect the writing efficiency in a
significant way. Figure \ref{fig:r=0.1}a shows the dependence of the
total losses on time for a given length in two cases: for exact
resonance (blue curve, dotted line) and with a detuning $r=0.1$ (red
curve, full line). The curves coincide to within 1.5\% over the full
range of $\tilde T^W$. Thus, despite the local differences in the
field-atom interaction in these two cases, the presence of a small
detuning does not actually change the properties of memory cell as a
whole. However when the detuning increases, the difference between
the curves increases (at $r=0.2$ it reaches 4.5\%, see Fig.
\ref{fig:r=0.1}b), but in the range of interest for $\tilde T^W$,
that is the one that allows minimization of the losses, the curves
are still close to each other (they agree within 1.5\%). Further
increase of the detuning distorts the profile of the losses further,
and the value of $\tilde T^W$ that provides minimum losses is
shifted (see Fig. \ref{fig:r=0.1}c,d).

\begin{figure}[h]
 \centering
 \includegraphics[height=40mm]{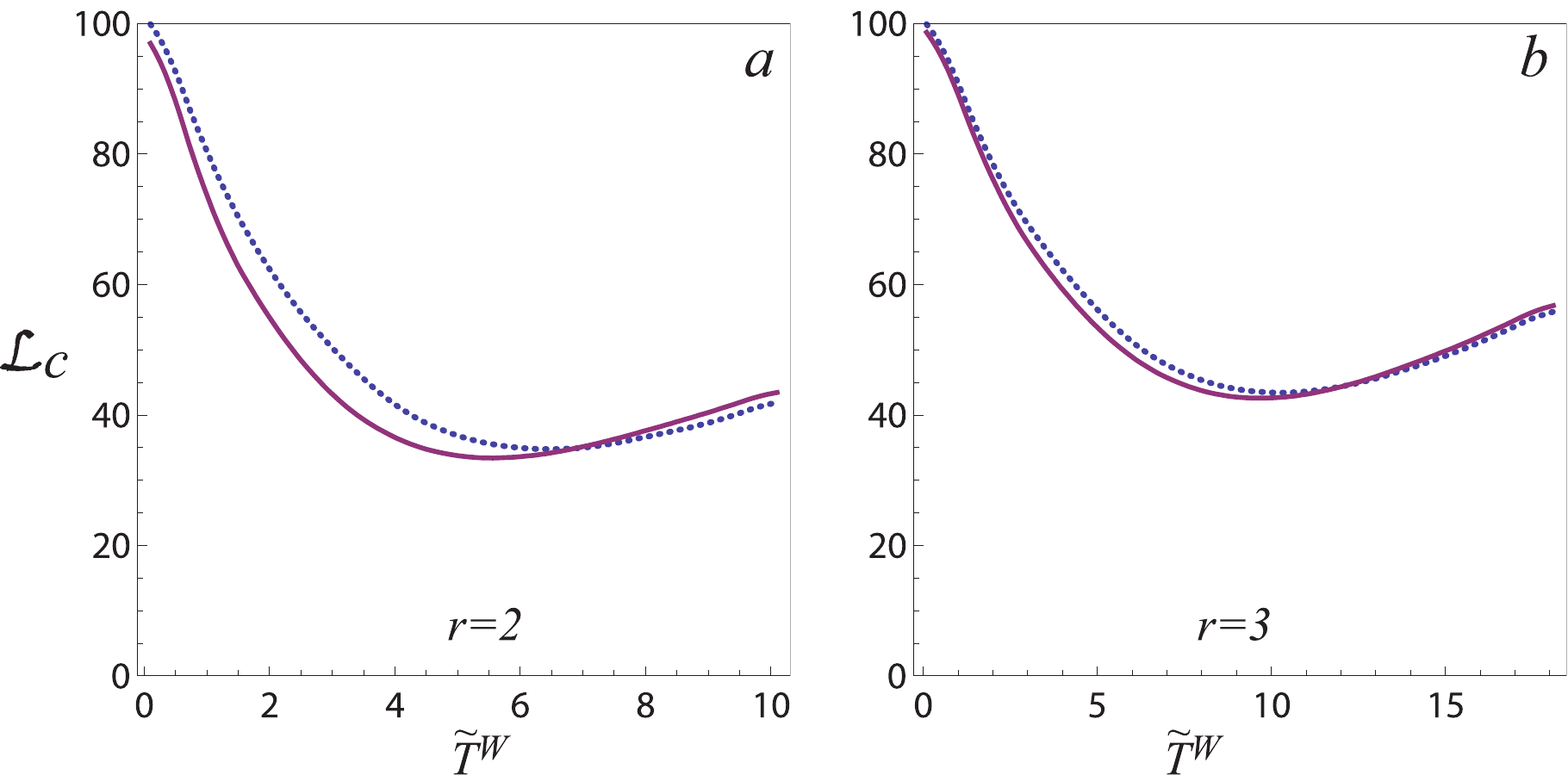}
   \caption{Comparison of the general calculations (blue curves, dotted lines) and calculations in the Raman limit (red curves, full lines).
   Writing process : relative total losses (in percent of the input field intensity) at the output
 of the medium as a function of $\tilde T^W$ for
  $\tilde L=10$ for (a)$r=2$ and (b) $r=3$. }
  \label{fig:Raman_Fig3}
\end{figure}

Let us now turn to the case of large detuning. One can see in Fig.
\ref{fig:Raman_Fig3} that for $r=3$ the profile of the total losses
calculated with the general formulas (\ref{3.1})-(\ref{3.2})
coincides well with the profile for the same quantity, calculated in
the Raman approximation with $r\gg 1$ (the difference is about
2.5\%, and it is less than 1\% at the minimum). For a smaller
detuning (for $r=2$) the difference between the curves increases up
to 7\% (about 3\% at the minimum): the calculation made in the Raman
case underestimates the minimum losses and shifts toward lower
values of $T^W$. Thus, we can conclude that for $r=3$ and higher the
Raman approach is applicable with good accuracy, but the general
solutions should used for lower values of the detuning.

For large enough detunings where the adiabatic limit is valid as well as our model, we can compare the results further. Our model predicts a storage
efficiency below the maximal available efficiency reached by the method proposed in reference \cite{Gorshkov2007}. In the third plot of Fig.
\ref{fig:losses} the efficiency is 65\%  for $Td\gamma=\tilde T^W \tilde L/2=30$ and $d=2400$. For the same parameters the adiabatic optimization
method converges to the maximal efficiency which is close to 100\%. This is due to the fact that we do not elaborate shaping of a control pulse, used
in reference \cite{Gorshkov2007}. Thus our method, even without control pulse shaping, is quite powerful in a non-adiabatic limit. Otherwise, the
numerical adiabatic optimization of the control pulse profile should be used to reach the maximal storage efficiency for the long pulses.

\section{Discussion of the read-out process \L{VIII}}

In Ref. \cite{Golubeva2011} the optimization of the retrieval
efficiency was based on the choice of the pulse duration providing
the minimum leakage for a given medium length. Here we will look for
a minimization of the total losses. Moreover, as was shown in Ref.
\cite{Gorshkov2007,Golubeva2011}, backward retrieval provides
significantly larger efficiency than forward retrieval, therefore we
will study the cases of forward and backward retrieval and compare
them. Using the result of Fig. \ref{fig:losses}a, we choose a value
for input signal duration of $\tilde T^W=5.5$ that provide  minimum
total losses for the writing process.

\begin{figure*}[h]
 \centering
 \includegraphics[height=40mm]{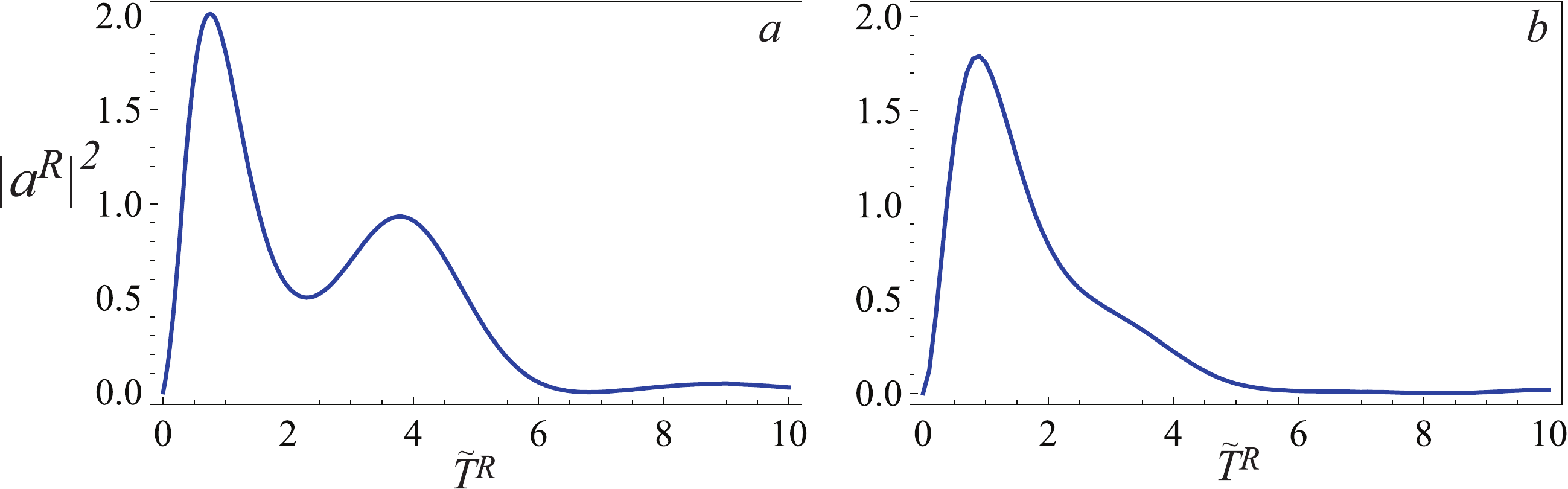}
 \caption{Reading process : field intensity $|a^R(\tilde T^R,\tilde L)|^2$ at the output of the medium for $\tilde L=10$
 and $r=0$ for backward retrieval with two optimization techniques: (a) total loss minimization ($\tilde T^W=5.5$)
 and (b) leakage minimization ($\tilde T^W=4.2$). }
  \label{a_R_resonant}
\end{figure*}

For this value of $\tilde T^W$ we calculate the intensity of the
retrieval field (normalized to the intensity of the input signal) as
a function of the reading time (see Fig. \ref{a_R_resonant}a). The
plot obtained from optimization based on leakage \cite{Golubeva2011}
is shown in Fig. \ref{a_R_resonant}b, in order to compare the
results. The retrieval efficiency is defined by:
\BE
{\cal E}(\vec q)=\frac{\int_0^{\tilde T^R}|a^R(\tilde t,\tilde z,\vec q)|^2 d \tilde t}{\int_0^{\tilde T^W}| a_{in}(\tilde t,\vec q)|^2 d \tilde
t\L{Eff}}\times100\%
\EE

In the case of optimization based on total losses, we find that the retrieval efficiency is equal to 88\% at $\tilde T^R=2\tilde T^W$. In view of the
writing efficiency (${\cal L}_c=11.2$\%), we see that using a reading time $\tilde T^R=2\tilde T^W$ we can restore almost all the information written
in the medium. When the leakage-based optimization is used, we find ${\cal E}=84\%$ at $\tilde T^R=3\tilde T^W$. This comparison clearly demonstrates
the benefits of the optimization based on the total losses.

 As for the temporal profiles of retrieval field, it is obvious
that in both cases they are very different from the input signal profile, which is a usual result in most memory processes.

\begin{figure*}[h]
 \centering
 \includegraphics[height=40mm]{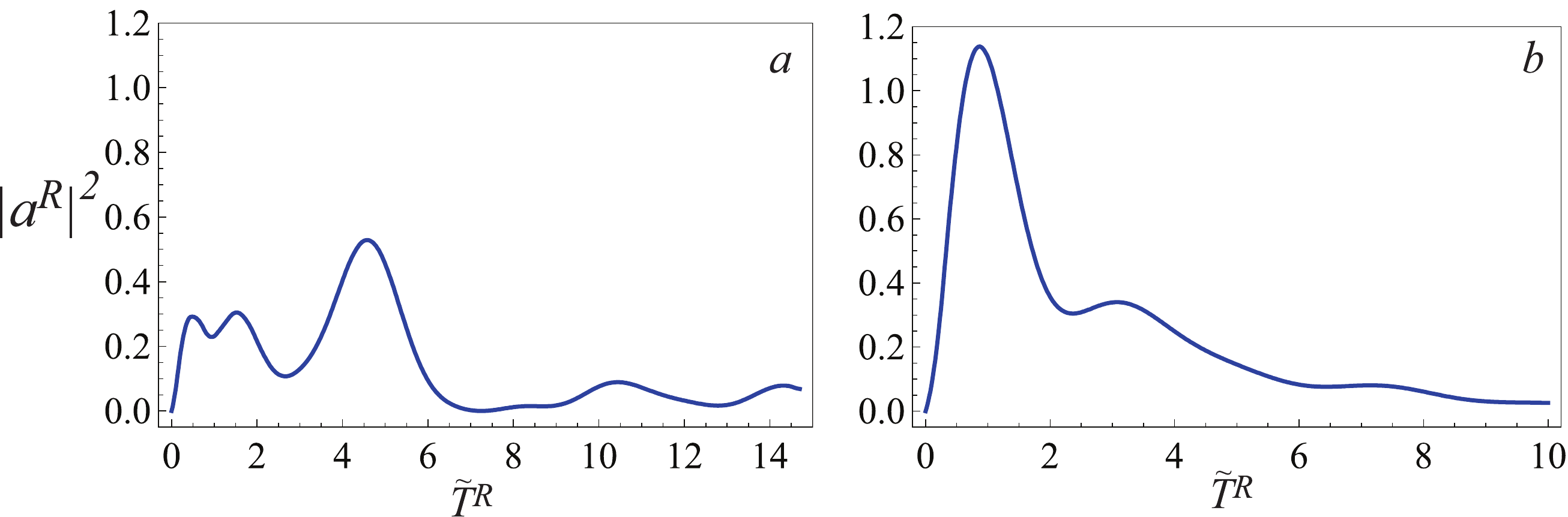}
 \caption{Reading process : field intensity $|a^R(\tilde t,\tilde L)|^2$ at the output of the medium for $\tilde L=10$,
  $\tilde T_W=3$ and $r=0.5$ for (a) forward and (b) backward propagating retrieval. }
  \label{a_R_non-resonant}
\end{figure*}

Let us now consider the result of the forward and backward retrieval for the non-resonant case(see Fig.~\ref{a_R_non-resonant} for $r=0.5$).

The efficiency is much lower for forward retrieval than for backward
retrieval one: at $\tilde T^R=10$ (i.e. $\tilde T^R\approx 3.3\tilde
T^W$) the efficiency of forward process is ${\cal E}=58\% $, while
the efficiency of backward process is ${\cal E}=85.6\% $. Moreover,
in the latter case less than 2\% of the available photons remain in
atomic ensemble (since the writing losses were 12.6\%). Comparison
with the resonant case shows that the presence of detuning slows
down the read-out.

\begin{figure}[h]
 \centering
 \includegraphics[height=40mm]{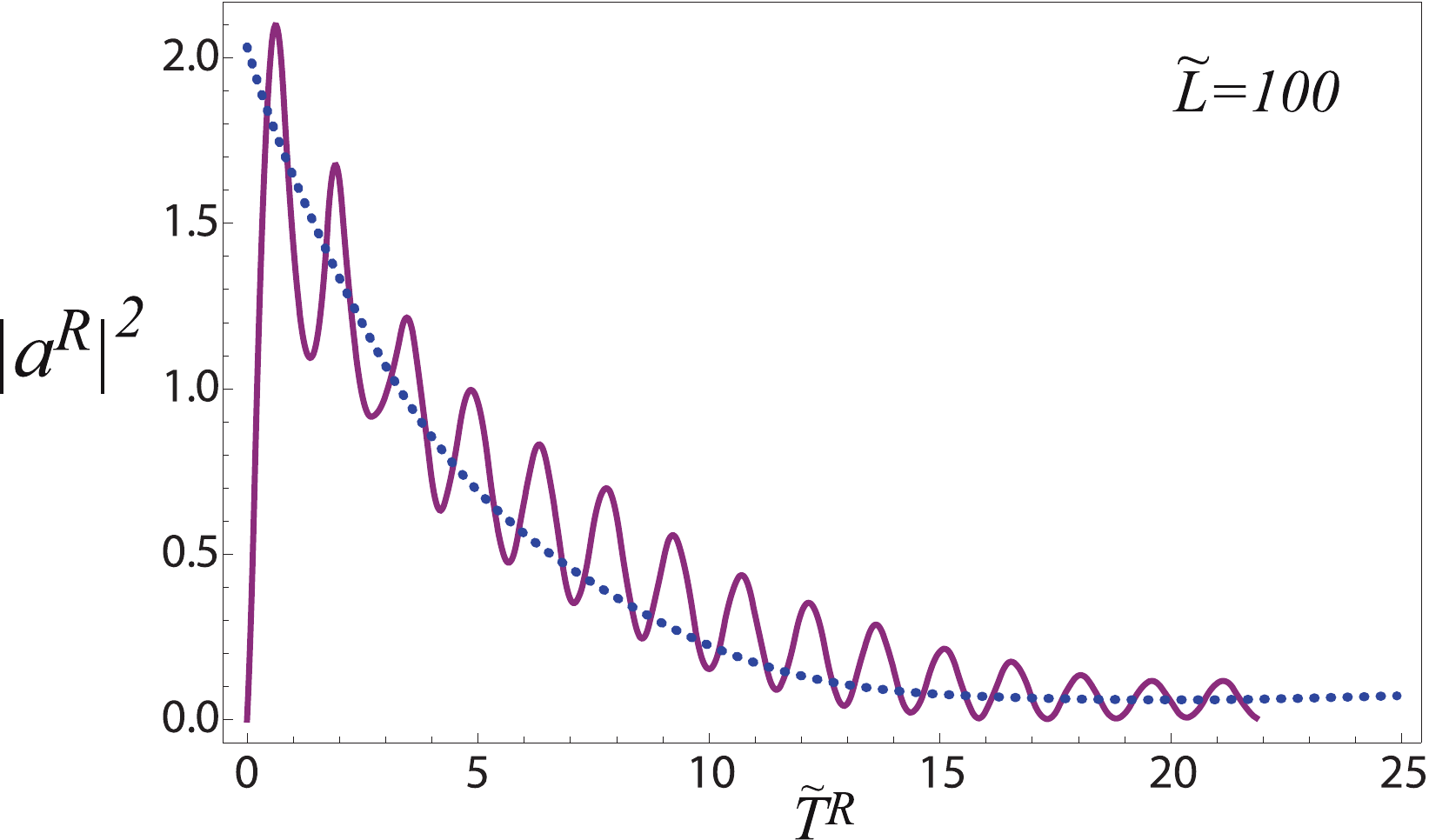}
 \caption{Reading process : field intensity $|a^R(\tilde t,\tilde L)|^2$ at the output of the medium for $\tilde L=100$,
  $\tilde T_W=4\pi$ and $r=2$ for backward retrieval calculated with exact formulas (red curve, full line)
  and within the Raman approximation (blue curve, dotted line). }
  \label{fig:a_R_back_2}
\end{figure}

Finally, we can show that even for a large detuning, the total efficiency can be large, at the condition that the medium length $\tilde L$ is large
enough. In Fig. \ref{fig:a_R_back_2} the red (oscillating) curve corresponds to the calculated read-out signal intensity for backward retrieval for
$\tilde L=100$, $r=2$, and for a pulse duration of the input field $\tilde T^W=4\pi$, which ensures a minimum in the total writing losses (writing
efficiency is $n^{eff}=95.6\% $). The oscillation period is equal to $4\pi r$, similar to the modulation observed in Fig.~\ref{fig:z=0}. If we
calculate the intensity of the retrieved signal in the Raman approximation, the overall shape of the curve remains the same, but the oscillations
disappear (blue curve, dotted line in Fig.~\ref{fig:a_R_back_2}). The retrieval efficiency calculated with these curves differ by 1.4\% (78.6\% for
the exact calculation and 77.2\% for the calculation in Raman approximation), so that the calculation in Raman limit can be a quite good estimation
for memory efficiency at $r=2$. Note, that here like for the curves in Fig. \ref{fig:z=0}, the magnitude of the oscillations will decrease with
increasing detuning.

In addition to high speed operation, our model includes transverse coordinates, as can be seen from Eq. (\ref{2.2}), and thus allows the treatment of
a variety of multimode fields, in the same way as in Ref.~\cite{Golubeva2011}. The storage of a multimode signal field is performed using a single
mode signal field, and the information is written in the atomic ensemble as a hologram. The quantum hologram process together with a Raman memory was
proposed for the first time by Sokolov et. al. in Ref.~\cite{Sokolov2010} and it was shown that the memory capacity is limited by diffraction, but
with the limitation depending on the direction of readout. Under forward readout the maximum number of the modes or picksels  $N$ which can be stored
in principle is given by the square of the Fresnel number $N\sim F_N^2$, where $F_N = S / (\lambda L)$. This expresses the condition that the output
pixel size should not exceed the transverse size of the memory cell. This rather loose limitation comes from the fact that most diffraction effects
are compensated between writing and readout with similar geometries. A similar compensation does not take place for the backward readout and as a
result the number of stored modes $N$ in this case can not exceed the Fresnel number.

\section{Conclusion}
In this article, we have presented a model for a high speed quantum
memory based on a three-level medium in the $\Lambda$-configuration.
This model relies on a full calculation of the fields and of the
atomic coherences as a function of time and space for arbitrary
frequency one-photon detuning, while keeping the two-photon
resonance. It allows to examine all the situations between
interactions resonant with the excited state and non resonant
interaction, which corresponds to a Raman transition. Our model
allows to identify the conditions in which the interaction can be
treated as a Raman transition. This corresponds to a normalized
detuning $r=\Delta/(2\Omega)>2$. For larger detunings, we have shown
that the Raman model gives accurate results.

In the near-resonant case, the interaction of the three-level atomic system with the weak signal field and the strong driving field turns out to be
more complicated than in the Raman case. As a matter of fact in contrast to the Raman process, where the upper atomic state is practically not
involved in the process, all three levels are populated, and this  leads ultimately to additional undesirable losses. We have shown that it is
possible to choose the parameters to make these losses very small. By controlling the Rabi oscillation one can preferentially populate the lower
state $|2\rangle$ and depopulate state $|3\rangle$. When the frequency detuning increases, the upper state population naturally decreases, which
contributes to the reduction of losses. Moreover, for large frequency detunings, the interaction between the atomic medium and the fields is weaker,
and a larger medium depth is necessary to reduce the leakage.

An important question is how to optimize the memory process  to obtain the highest possible quantum efficiency.  In the literature two approaches are
mainly discussed. The first one is based on the choice of the  optimal time-shape for the signal pulse, using for this the eigenfunctions of the
integral operators specific of the considered memory \cite{Nunn2007}. It was demonstrated that if the functions possess some symmetry, the efficiency
can be improved. In our case with very short pulses, there is no such symmetry and this optimization turns out to be impossible. Another optimization
method was proposed by Gorshkov et. al. in Ref.~\cite{Gorshkov2007}, based on a study of the driving pulse shape. However, the search procedure for
the optimal shape in this article is based on the adiabatic or Raman approximations, which are not generally applicable in the case of very short
pulses. Non-adiabatic pulse shape optimization was also developed for the resonant case \cite{Gorshkov2008} extending the validity of the method. In
our model, the optimization is based on the minimization of the losses, which come from the leakage and from the population of the upper atomic
level, and allows identifying the optimal combination of signal pulse duration and optical depth. We demonstrate that high memory efficiencies can be
achieved by this method for very short pulses whatever the value of the detuning.

\section{Acknowledgements} The study was performed within the
framework of the Russian-French Cooperation Program "Lasers and
Advanced Optical Information Technologies", of the European Project
HIDEAS (grant No. 221906). O.M. acknowledges the support of Ile de
France programme IFRAF.  The study was also supported by RFBR (grant
No. 08-02-92504).

\appendix

\section{General solutions for the main equations \L{A}}
Eqs.~(\ref{2.19})-(\ref{2.21}) can be solved in the general form by
using the Laplace domain. Some details of the formal procedures are
discussed  in App.~\ref{B} and now we start with the solutions
themselves in the explicit form. For our analysis in this article we
do not need full information about the solutions, nevertheless one
can find it below. Here and everywhere in article under
consideration of the solution we use dimensionless co-ordinates
$\tilde t$ and $\tilde z$ given by
\BY
&&\tilde t=\Omega t,\qquad \tilde z=\frac{2g^2N}{\Omega}z.
\EY
Let us write the general solutions in the form under the arbitrary initial and boundary conditions
\BY
\hat a(\tilde t,\tilde z;\vec q)=&&\int_0^{\tilde t}d\tilde t^\prime \hat a_{in}(\tilde t-\tilde t^\prime;\vec q)G_{aa}(\tilde t^\prime,\tilde
z)\nn\\
&&-\frac{1}{2p}\int_0^{\tilde z}d\tilde z^\prime \hat
b(0,\tilde z-\tilde z^\prime;\vec q)G_{ba}(\tilde t,\tilde z^\prime)\nn\\
&&-\frac{1}{2p}\int_0^{\tilde z}d\tilde z^\prime \hat c(0,\tilde z-\tilde z^\prime;\vec q)
G_{ca}(\tilde t,\tilde z^\prime),\L{A1}\\
\hat b(\tilde t,\tilde z;\vec q)=&&-p\;\int_0^{\tilde t}d\tilde t^\prime \hat a_{in}(\tilde t-\tilde t^\prime;\vec
q)G_{ab}(t^\prime,z)\nn\\
&&+\frac{1}{2}\int_0^{\tilde z}d\tilde z^\prime \hat
b(0,\tilde z-\tilde z^\prime;\vec q)G_{bb}(\tilde t,\tilde z^\prime)\nn\\
&&+\frac{1}{2}\int_0^{\tilde z}d\tilde z^\prime \hat c(0,\tilde z-\tilde z^\prime;\vec q)G_{cb}(\tilde t,\tilde z^\prime),\L{A2}\\
\hat c(\tilde t,\tilde z;\vec q)=&&p\int_0^{\tilde t}d\tilde t^\prime \hat a_{in}(\tilde t-\tilde t^\prime;\vec q)G_{ac}(\tilde t^\prime,\tilde
z)\nn\\
&&+\frac{1}{2}\int_0^{\tilde z}d\tilde z^\prime \hat b(0,\tilde z-\tilde z^\prime;\vec
q)G_{bc}(\tilde t,\tilde z^\prime)\nn\\
&&+\frac{1}{2}\int_0^{\tilde z}d\tilde z^\prime \hat c(0,\tilde z-\tilde z^\prime;\vec q)G_{cc}(\tilde t,\tilde z^\prime),\L{A3}
\EY
where kernels $G_{ik}(\tilde t,\tilde z)$ are bilinear combinations of the expressions depending on the n-th Bessel functions of the first kind
denoted by $J_{n}$
\BY
&&f(\tilde t, \tilde z;r)=\delta(\tilde t)-e^{\ds -i\(\sqrt{1+r^2} +r\)\tilde t}\sqrt{\frac{(1+r)\tilde z}{4\tilde t}}\nn\\
&&\qquad\qquad\times J_1\(\sqrt{(1+r)\tilde t\tilde z}\)\Theta(\tilde t),\L{A4}\\
&&f_1(\tilde t, \tilde z;r)=e^{\ds-i\(\sqrt{1+r^2} +r\)\tilde t}\;\sqrt{\frac{4(1+r)\tilde t}{ \tilde z}}\nn\\
&&\qquad\qquad\times J_1\(\sqrt{(1+r)\tilde t\tilde z}\)\Theta(\tilde t),\L{A6}\\
&&f_0(\tilde t, \tilde z;r)=e^{\ds-i\(\sqrt{1+r^2} +r\)\tilde t}\nn\\
&&\qquad\qquad\times J_0\(\sqrt{(1+r)\tilde t\tilde z}\)\Theta(\tilde t).\qquad\L{A8}
\EY
The kernels in Eq.~(\ref{A1}) read
\BY
&&G_{aa}(\tilde t,\tilde z)=\[f(r)\ast f^\ast(-r)\](\tilde t,\tilde z),\L{A10}\\
&&G_{ba}(\tilde t,\tilde z)=[f_0(r)\ast f_0^\ast(-r)](\tilde t,\tilde z),\L{A11}\\
&&G_{ca}(\tilde t,\tilde z)=\frac{1+r}{2}[f_0(r)\ast f^\ast(-r)](\tilde t,\tilde z)\nn\\
&&\qquad\qquad+\frac{1-r}{2}[f(r)\ast f_0^\ast(-r)](\tilde t,\tilde z).\L{A12}
\EY
We denote the t convolution of two arbitrary functions $X(\tilde t,\tilde z;r)$ and $Y(\tilde t,\tilde z;r)$  as
\BY
&&[X(r)\ast Y^\ast(-r)](\tilde t,\tilde z)=\nn\\
&&\qquad\qquad=\int_0^{\tilde t}d\tilde t^\prime X(\tilde t-\tilde t^\prime,\tilde z;r)Y^\ast(\tilde t^\prime,\tilde z;-r).\qquad\L{A14}
\EY
The  parameter $r$ is the dimensionless frequency detuning $r={\Delta}/({2\Omega})$.
\\
The other kernels read
\BY
&&G_{ab}(\tilde t,\tilde z)=[f_0(r)\ast f_0^\ast(-r)](\tilde t,\tilde z),\L{A15}\\
&&G_{bb}(\tilde t,\tilde z)=2\;\delta(\tilde z)F_1(\tilde t)+ [f_1(r)\ast f_1^\ast(-r)](\tilde t,\tilde z),\L{A16}\\
&&G_{cb}(\tilde t,\tilde z)=2\;\delta(\tilde z)F_2(\tilde t)+\frac{1+r}{2}[f_1(r)\ast f_0^\ast(-r)](\tilde t,\tilde z)\nn\\
&&\qquad\qquad+\frac{1-r}{2}[f_0(r)\ast f_1^\ast(-r)](\tilde t,\tilde z),\L{A17}
\EY
and
\BY
&&G_{ac}(\tilde t,\tilde z)=\frac{1+r}{2}[f_0(r)\ast f^\ast(-r)](\tilde t,\tilde z)\nn\\
&&\qquad\qquad+\frac{1-r}{2}[f(r)\ast f_0^\ast(-r)](\tilde t,\tilde z),\L{A18}\\
&&G_{bc}(\tilde t,\tilde z)=2\;\delta(\tilde z)F_2(\tilde t)-\frac{1+r}{2}[f_1(r)\ast
f_0^\ast(-r)](\tilde t,\tilde z)\nn\\
&&\qquad\qquad-\frac{1-r}{2}[f_0(r)\ast f_1^\ast(-r)](\tilde t,\tilde z)],\L{A19}\\
&&G_{cc}(\tilde t,\tilde z)=2\;\delta(\tilde z)F_3(\tilde t)-\frac{(1+r)^2}{4}[f_1(r)\ast f^\ast(-r)](\tilde t,\tilde
z)\nn\\
&&\qquad\qquad-\frac{(1-r)^2}{4}[f(r)\ast f_1^\ast(-r)](\tilde t,\tilde z)\nn\\
&&\qquad\qquad-\frac{1-r^2}{2}[f_0(r)\ast f_0^\ast(-r)](\tilde t,\tilde z).\L{A20}
\EY
In these formulas the time dependent factors $F(\tilde t)$ read
\BY
&&F_1(\tilde t)=\[\cos\(\sqrt{1+r^2} \;\tilde t\)\right.\nn\\
&&\qquad\qquad\left.+\frac{ir}{\sqrt{1+r^2} }\sin\(\sqrt{1+r^2} \;\tilde t\)\]\; e^{-ir\tilde t},\L{A21}\\
&&F_2(\tilde t)=\frac{1}{\sqrt{1+r^2}}\sin\(\sqrt{1+r^2} \;\tilde t\)\;e^{-ir\tilde t},\L{A22}\\
&&F_3(\tilde t)=\[\cos\(\sqrt{1+r^2} \;\tilde t\)\nn\right.\\
&&\qquad\qquad\left.-\frac{ir}{\sqrt{1+r^2} }\sin\(\sqrt{1+r^2} \;\tilde t\)\]e^{-ir\tilde t}\L{A24}.
\EY

\section{Solutions using the Laplace transform  \L{B}}

 In order to solve Eqs.~(\ref{2.19})-(\ref{2.21}),  we rewrite them in the Laplace domain
\BY
&& \frac{d }{d z}\hat a_s(z;\vec q) =- g \sqrt N\hat c_{s}(z;\vec q)\L{B1}\\
 &&-\hat c(z,0;\vec q)+(s+i\Delta)\hat c_s(z;\vec q)= g\sqrt N  \hat a_s(z;\vec q)+
 \Omega \hat b_s(z;\vec q),\nn\\
 &&\L{B2}\\
 && -\hat b(z,0;\vec q)+s\hat b_s(z;\vec q)= - \Omega\hat c_s(z;\vec q) \L{B3}.
\EY
From this we can write the closed differential equation for the field amplitude $\hat a_s(z;\vec q)$ in the form
 \BY
&&\frac{d\hat a_s(z;\vec q)}{dz}=-\Gamma_s\hat a_s(z;\vec q)-g\sqrt N\hat \alpha_s(z;\vec q).\L{B4}
\EY
Here the coefficient $\Gamma_s$ determines a rate of escape of the amplitude along the z-axis and reads
 \BY
&&\Gamma_s=\frac{g^2N}{2}\(\frac{\mu}{s+i \mu\tilde\Omega}+\frac{\nu}{s-i \nu\tilde\Omega}\),\L{B5}
\EY
where the following notations are introduced
 \BY
&&\tilde \Omega=\Omega\sqrt{1+r^2},\qquad\mu=1+r,\qquad\nu=1-r.\qquad\L{B5}
\EY
The inhomogeneous  term on the right  in Eq.~(\ref{B4}) is
determined by the initial conditions for the medium state and given
by
\BY
&& \hat \alpha_s(z;\vec q)=\frac{1}{s(s+i\Delta)+\Omega^2}\[\Omega\hat b(0,z;\vec q)+s\hat c(0,z;\vec q)\].\qquad\L{B7}
\EY
A solution of Eq.~(\ref{B7}) reads
 \BY
&&\hat a_s(z;\vec q)=\hat a_s(0;\vec q)e^{-\Gamma_sz}-g\sqrt N\int\limits_0^z dz^\prime \hat \alpha_s(z^\prime;\vec
q)e^{-\Gamma_s(z-z^\prime)}.\nn\\
&&\L{B8}
\EY
From Eqs.~(\ref{B2})-(\ref{B3}) one can obtain
\BY
&&\hat c_s(z;\vec q)= g\sqrt N\frac{s}{s(s+i\Delta)+\Omega^2}\hat a_s  (z;\vec q)+\hat\alpha_s(z;\vec q),\qquad\\
&&\hat b_s(z;\vec q)= \frac{1}{s}\[\hat b(0,z;\vec q)-\Omega\hat c_s(z;\vec q)\].\L{10}
\EY
After the inverse Laplace transformation in Eqs.~(\ref{B8})-(\ref{10}) one can obtain   all the solutions  in the form
\BY
&& \hat a(t,z;\vec q)=\int_0^t dt^\prime \hat a_{in}(t-t^\prime;\vec q)\; D(t^\prime, z)\L{B10}\\
&&\qquad -g\sqrt N\int_0^t dt^\prime\int_0^z
dz^\prime \hat \alpha(t-t^\prime, z-z^\prime;\vec q)\; D(t^\prime, z^\prime),\nn\\
&& \hat c(t,z;\vec q)= \nn\\
&&\qquad g\sqrt N\int_0^t dt^\prime F_3(t-t^\prime)\; \hat a(t^\prime,z;\vec q)+\hat \alpha(t,z;\vec q),\L{B11}\\
 &&\hat b(t, z;\vec q)=\nn\\
&&\qquad -g\sqrt N\int_0^t dt^\prime F_2(t-t^\prime) \; \hat a(t^\prime,z;\vec q)+ \hat \beta(z,t;\vec q),\L{B12}
 \EY
where the terms $\hat \alpha(t,z;\vec q)$ and $\hat \beta(z,t;\vec q)$ are expressed via the initial conditions for the medium coherences in the form
 \BY
 && \hat \alpha(t,z;\vec q)=F_3(t)\;
 \hat c(0,z;\vec q)+F_2(t)\;\hat b(0,z;\vec q),\L{B16}\\
 &&\hat \beta(z,t;\vec q)=-F_2(t)\;\hat c(0,z;\vec q)+
F_1(t)\;\hat b(0,z;\vec q),\L{B17}
\EY
and the kernel $D(t,z)$ is simply proportional to $G_{aa}$ (see (\ref{A8})) and is given by
\BY
&& D(t,z)=\Omega G_{aa}(\tilde t,\tilde z).
\EY

After some simple transformations one can obtain the solution in the form (\ref{A1})-(\ref{A3}). At the same time for the numerical computation it is
possible to use Eqs.~(\ref{B10})-(\ref{B12}).

\end{document}